\documentclass{article}
\usepackage[utf8]{inputenc}
\usepackage{graphicx}
\usepackage{color}
\usepackage{amsfonts}
\usepackage{tabularx}
\usepackage{booktabs}
\usepackage{adjustbox} 
\usepackage{authblk}
\usepackage{longtable}
\usepackage{amsmath}
\usepackage{relsize}
\usepackage{tabularx} 
\usepackage{bbm}
\usepackage[a4paper, margin=1in]{geometry}

\newcommand\ddfrac[2]{\frac{\displaystyle #1}{\displaystyle #2}}

\normalsize  

\title{RISE: Two-Stage Rank-Based Identification of High-Dimensional Surrogate Markers Applied to Vaccinology}

\author[1,2]{Arthur Hughes}
\author[3]{Layla Parast}
\author[1,2,4]{Rodolphe Thiébaut}
\author[1,2]{Boris P. Hejblum}

\affil[1]{INSERM Bordeaux Population Health Research Center, INRIA SISTM, University of Bordeaux, F-33000 Bordeaux, France}
\affil[2]{Vaccine Research Institute, F-94000 Créteil, France}
\affil[3]{Department of Statistics and Data Science, University of Texas at Austin, Austin, TX 78712, United States of America}
\affil[4]{Centre Hospitalier Universitaire de Bordeaux, Service d’Information Médicale, INSERM, Bordeaux F-33000, France}

\date{} 

\begin{document}

\maketitle

\begin{abstract}
In vaccine trials with long-term participant follow-up, it is of great importance to identify surrogate markers that accurately infer long-term immune responses. These markers offer practical advantages such as providing early, indirect evidence of vaccine efficacy, and can accelerate vaccine development while identifying potential biomarkers. High-throughput technologies like RNA-sequencing have emerged as promising tools for understanding complex biological systems and informing new treatment strategies. However, these data are high-dimensional, presenting unique statistical challenges for existing surrogate marker identification methods. We introduce Rank-based Identification of high-dimensional SurrogatE Markers (RISE), a novel approach designed for small sample, high-dimensional settings typical in modern vaccine experiments. RISE employs a non-parametric univariate test to screen variables for promising candidates, followed by surrogate evaluation on independent data. Our simulation studies demonstrate RISE's desirable properties, including type one error rate control and empirical power under various conditions. Applying RISE to a clinical trial for inactivated influenza vaccination, we sought to identify genes whose expression could serve as a surrogate for the induced immune response. This analysis revealed a signature of genes appearing to function as a reasonable surrogate for the neutralising antibody response. Pathways related to innate antiviral signalling and interferon stimulation were strongly represented in this derived surrogate, providing a clear immunological interpretation.
\end{abstract}

\vspace{1em} 

\newpage

\section{Introduction}\label{sec1}

The goal of a randomised clinical trial is typically to evaluate the efficacy of a treatment on a primary outcome. However, measuring this primary outcome can be time-consuming, costly, impractical, or unethical. Consequently, there is significant interest in identifying and validating surrogate markers that can accurately infer treatment effects on the primary outcome without its direct observation \cite{Christensen_2024}.

The identification of surrogate markers is especially important in the context of vaccine clinical trials. In public health emergencies, such as the COVID-19 pandemic, validated surrogates enable accelerated vaccine development by providing evidence for candidate vaccine selection in early stage trials \cite{Gilbert2024}. Furthermore, surrogate markers may allow for validation of new-generation vaccines where an efficacy trial would not be ethical or feasible \cite{Plotkin_2010}.

High-throughput technologies have emerged as promising candidates to better inform effective vaccine design \cite{Weiner2015}. A prime example is transcriptomic data, which describe gene expression \--- a dynamic process through which fixed information encoded in DNA is transformed into proteins which then in turn shape phenotypes. The gene expression response to vaccination is highly upstream, with changes observed in the first days following vaccination which mainly capture innate immune responses \cite{Hou2017}. Traditional downstream immunological outcomes, however, such as antigen-specific antibody titres or T-cell responses, become fully established over weeks and months \cite{Furman2015}. Therefore, a subset of genes whose expression serves as a surrogate for a vaccine's immunogenicity would have significant utility. For example, such a surrogate could be used to down-select vaccine strategies from a large number of candidates in adaptive clinical trials that rely on rapidly measurable endpoints \cite{Liu2021}.
Additionally, studies have previously derived early gene expression signatures which are correlated with, or predictive of, immune responses to vaccination \cite{Querec2008, Kazmin2017, Nakaya2011}. However, a good predictor or correlate does not necessarily make a good surrogate \cite{Baker2003}\---whether these gene expression markers (either individually or in combination) could serve as reliable surrogates for the vaccine response remains to be seen. 

The statistical methods for evaluating surrogate markers have evolved considerably over the past four decades \cite{Elliott2023}. Prentice's seminal paper defined a surrogate marker as one for which any test of the treatment's effect on the surrogate is also a valid test of the treatment's effect on the primary endpoint \cite{Prentice1989}. In line with this definition, a substantial body of literature has since focused on assessing measures such as the proportion of the treatment effect on the primary response that is captured by the surrogate marker (PTE), and the ratio of the treatment's effect on the primary response to that on the surrogate (RE) \cite{LIN1997,Wang2002,Li2001,Roberts2023,Buyse2000,Molenberghs_2002}. However, the available methods often rely on parametric model specifications; if the true underlying model is unknown, any conclusions drawn about the surrogate markers may be invalid. These parametric assumptions are particularly difficult to verify when sample sizes are small. Moreover, the few non-parametric alternatives available depend on kernel smoothing techniques, which tend to perform poorly without a large sample \cite{Parast_2015,Agniel2020}. Consequently, current methods are generally not well suited to small sample settings. Lastly, surrogate evaluation becomes even more complex in settings involving multiple surrogate markers, especially when dealing with a high-dimensional set of candidate surrogates. While some existing approaches allow for the assessment of the overall strength of a collection of candidate surrogates \cite{Agniel2020,Zhou_2022}, they do not provide a means to screen high-dimensional data to identify a subset of markers that effectively capture a relevant proportion of the treatment's effect on the response.

These limitations make existing methodology difficult to apply in practice to vaccine trials, where the model relating the primary outcome to the surrogates is complex and unknown, the sample size is typically small, and the candidate surrogates may be high-dimensional. Motivated by the application to vaccine development, we extend a recent approach by Parast \textit{et al.} (2024) \cite{Parast2024}, which is a fully non-parametric rank-based approach, to the multiple marker setting.  We propose \textit{Rank-based Identification of high-dimensional SurrogatE markers} (RISE) \--- a two step approach, which first screens a set of high-dimensional variables, and then evaluates their overall strength as a surrogate. We apply this approach to public data on seasonal inactivated influenza vaccination, seeking to identify gene expression surrogate markers of the vaccine immunogenicity.   

\section{Methods}\label{sec2}

\subsection{Notation}

Let $n$ denote the total sample size, which is assumed to be small. Let $Y$ denote the primary outcome, $A \in \{1,0\}$ denote a binary vaccine indicator (e.g., vaccine or placebo), and $\boldsymbol{S} = \left(S_{1},\dots,S_{p}\right)$ denote a set of candidate surrogates, where we may have $p >> n$. Without loss of generality, we assume higher values of $Y$ and $S_j$ are ``better'', with $j\in\{1,\dots,p\}$. We adopt counterfactual notations, where each individual has a set of potential outcomes $\left[ Y^{1}, Y^{0}, \boldsymbol{S}^{1}, \boldsymbol{S}^{0}\right]$. Here, $Y^{a}$ and $\boldsymbol{S}^{a}$ represent the values of the primary outcome and surrogate markers, respectively, if the treatment had been, potentially counter to fact, set to $A = a$. The observed data consist of $n_{1}$ independent, identically distributed (i.i.d.) copies of $Y^{1}, \boldsymbol{S}^{1}$ for individuals in the treatment group and $n_{0}$ i.i.d. copies of $Y^{0}, \boldsymbol{S}^{0}$ for individuals in the control group, with $n = n_0 + n_1$.

\subsection{Existing Rank-based Approach for a Single Surrogate}

We begin by summarising an approach proposed by Parast \textit{et al.} (2024) to evaluate a single surrogate in the small-sample setting \cite{Parast2024}. This method focuses on identifying \textit{trial-level surrogates}, which are markers for which the average treatment effect on the surrogate is predictive of the average treatment effect on the primary response. We emphasise that it does not, however, guarantee the identification of \textit{individual-level surrogates}, which predict treatment effects at the individual level \cite{Gabriel2016}. Going forward, the reader may assume any mention of surrogate validity refers to that on the trial-level and not on the individual-level.  

Motivated by Prentice's definition of a surrogate, this approach aims to identify a single surrogate candidate $S_{j}$ as valid if a test for a treatment effect based on the surrogate is a valid test for the treatment effect based on the primary outcome. Let 

$$U_{Y} = \mathbb{P}(Y^{1} > Y^{0}) + \frac{1}{2}\mathbb{P}(Y^{1} = Y^{0})$$

$$U_{S_{j}} = \mathbb{P}(S_{j}^{1} > S_{j}^{0}) + \frac{1}{2}\mathbb{P}(S_{j}^{1} = S_{j}^{0}),$$

\noindent where $U_{Y}$ is simply a measure of the treatment effect on $Y$, and $U_{Y} \in (0.5,1\rbrack$ indicates a positive treatment effect on $Y$, $U_{Y} \in \lbrack0,0.5)$ indicates a negative effect on $Y$, and $U_{Y} = 0.5$ indicates no effect. Similarly, $U_{S_{j}}$ quantifies the treatment effect on $S_j$. The general idea behind this approach is that the closer $U_Y$ and $U_{S_{j}}$ are to each other, the more $S_{j}$ captures the average treatment effect on $Y$ and thus, is a stronger surrogate marker for $Y$. The strength of the surrogate is quantified by the difference

$$\delta_{j} = U_{Y} - U_{S_{j}}$$

\noindent such that the closer $\delta_{j}$ is to $0$, the stronger $S_{j}$ is as a surrogate for $Y$. One could then consider $S_{j}$ to be a valid surrogate if $\delta_{j}$ is bounded by some pre-specified upper bound $\epsilon$. This is formalised through a non-inferiority test:

\begin{equation}
H_{0}^{} : \delta_{j} \geq \epsilon \hspace{0.5cm} \text{versus} \hspace{0.5cm} H_{1}^{} : \delta_{j} < \epsilon \label{noninf}
\end{equation}

\noindent where failure to reject $H_{0}^{}$ reflects a poor surrogate, and rejection of $H_{0}^{}$ reflects a valid surrogate. The quantities $U_{Y}$ and $U_{S_{j}}$ can be estimated as
$$\widehat{U}_{Y} = (n_{1}n_{0})^{-1} \mathlarger{\sum}\limits_{i =1}^{n_{1}}\mathlarger{\sum}\limits_{k =1}^{n_{0}}G(Y_{i1}, Y_{k0})\qquad\text{and}\qquad\widehat{U}_{S_{j}} = (n_{1}n_{0})^{-1} \mathlarger{\sum}\limits_{i =1}^{n_{1}}\mathlarger{\sum}\limits_{k =1}^{n_{0}}G(S_{ji1}, S_{jk0})$$
where 
$$G(A,B) = \begin{cases}
  1 , \quad \mbox{if}\quad  A> B\\
  \frac{1}{2}, \quad \mbox{if} \quad A=B\\
  0, \quad \mbox{if}\quad  B<A
\end{cases}$$
and $Y_{ia}$ and $S_{jia}$ denote the observed values of the primary response and $j$\textsuperscript{th} surrogate for individuals $i$ such that $A_{i} = a$. Here, $\widehat{U}_Y$ is simply the rank-based Mann-Whitney U-statistic examining the difference in $Y$ between the two  groups and similarly for $\widehat{U}_{S_j}$, where $E(\widehat{U}_Y) = U_Y$ and $E(\widehat{U}_{S_j}) = U_{S_j}$ \cite{Mann1947}.

Parast \textit{et al.} (2024) shows that $U_S > \frac{1}{2} \implies U_Y > \frac{1}{2}$ under the following conditions:
\begin{align*}
(C1) \quad & P(Y^a > y \mid S^a = s) \text{ increases in } s \\
(C2) \quad & P(S^1 > s) \geq P(S^0 > s) \quad \forall s \\
(C3) \quad & P(Y^1 > y \mid S^1 = s) \geq P(Y^0 > y \mid S^0 = s) \quad \forall s
\end{align*}

Together, these conditions establish that treatment has a non-negative effect on the surrogate (C2), that the surrogate has a non-negative association with the outcome (C1), and that, in the case where the surrogate does not capture all the treatment effect, the residual treatment effect on the outcome after conditioning on the surrogate is non-negative (C3). Under these assumptions, we avoid a so-called \textit{surrogate paradox} situation, where the treatment effect on the surrogate is positive, the surrogate is positively associated with the primary outcome, but the treatment effect on the primary outcome is negative. We note that these conditions alone do not establish surrogate validity ---that is addressed by $\delta_{j}$ and the testing procedure described below.

Then, for a given $S_{j}$, one can calculate $\widehat{\delta_{j}} = \widehat{U}_{Y} - \widehat{U}_{S_j}$. A closed-form expression for the standard deviation of $\widehat{\delta_{j}}$, denoted $\widehat{\sigma}_{\delta_j}$, is given in Parast \textit{et al.} (2024) and is based on theory for correlated U-statistics \cite{DeLong1988}. Let $\Phi^{-1}(.)$ denote the inverse cumulative distribution function of the standard normal distribution $\mathcal{N}(0,1)$. Then, given a nominal significance level $\alpha$, a one sided confidence interval for $\delta_j$ can be obtained as
$$\left[-1,\, \widehat{\delta}_j + \Phi^{-1}(1- \alpha) \widehat{\sigma}_{\delta_j}\right].$$

It can be shown that $\widehat{\delta}_j \sim N(\delta_j, \widehat{\sigma}_{\delta_j})$, and thus, taking the boundary of the null hypothesis in (\ref{noninf}), $\delta_j = \epsilon$, the p-value for testing $H_{0}$ can be calculated as $p_{j} = P\left(Z<\widehat{\delta}_j\right)$ where $Z \sim N\left(\epsilon, \widehat{\sigma}_{\delta_j}\right)$. A p-value $p_{j}<\alpha$ therefore corresponds exactly to the case where the upper confidence interval of $\widehat{\delta}_j$ is less than $\epsilon$.  

Of course, the choice of $\epsilon$ is subject to debate. One could choose a fixed low value of $\epsilon$ \textit{a priori} based on context or clinical guidance, but in the absence of prior knowledge, one can instead choose $\epsilon$ adaptively, as described in Parast, 2024 \cite{Parast2024}. Specifically, if the estimated treatment effect is $\widehat{U}_{Y}$, the significance level $\alpha$ and the desired power to detect a treatment effect based upon the candidate surrogate $S_{j}$ is $(1-\beta)$, one may select $\epsilon$ as: 
\begin{equation}
    \epsilon = \max\left\{0, \widehat{u}_{Y} - u^{*}_{\alpha, \beta}\right\}
\end{equation}
where 
\begin{equation*}
    u^{*}_{\alpha, \beta} = \frac{1}{2} - \sqrt{\frac{n_{0} + n_{1} + 1}{12n_{0}n_{1}}} \left[ \Phi^{-1}(\beta) - \Phi^{-1}(1-\alpha) \right].
\end{equation*}

The motivation behind this approach is that $u^{*}_{\alpha, \beta}$ (obtained with some algebra) is the value of $U_{S_j}$ where we would have exactly $\left(1-\beta\right)$ power to detect a treatment effect on $S_j$. Defining $\epsilon$ as shown in \eqref{eps_def} implies that we are willing to consider $S_j$ as a surrogate for $Y$ even if it is not as ``good'' as $Y$ in terms of capturing the treatment effect, as long as it has a certain minimum power which we specify. That is, our threshold is $\widehat{u}_{Y} - u^{*}_{\alpha, \beta}$ and is determined by our desired power. 

\subsection{Extensions to the Rank-Based Approach}

\paragraph*{\textbf{Paired-Sample Setting} \\ } 

In certain contexts, investigators may encounter paired data, such as pre‐ and post‐treatment measurements on the same individuals, or observations on matched individuals receiving different treatments. The test described above presupposes independence between treatment groups, an assumption that does not hold in this setting. Hence, we propose an extension to accommodate the paired-sample design, adequately accounting for the within-unit correlation.

In this framework, the observed data comprise $n$ independent and identically distributed observations of paired units $i$, with primary response vectors given by $\boldsymbol{Y}_i = \left(Y_i^1, Y_i^0\right)^{T}$ and surrogate candidate vectors by $\boldsymbol{S}_i = \left(S_i^1, S_i^0\right)^{T}$. The treatment effects $U_Y$ and $U_S$ are defined as above, and may be estimated by comparing each unit’s outcomes under the two treatments,

$$\widehat{U}_Y = \frac{1}{n}\sum_{i=1}^{n} G(Y_i^1, Y_i^0) \qquad \text{and} \qquad \widehat{U}_S = \frac{1}{n}\sum_{i=1}^{n} G(S_i^1, S_i^0).$$

The variance derivations of $U_Y, U_{S},$ and $\delta$ associated with this extension are described in the Supporting Information.

\paragraph*{\textbf{Two one-sided test procedure}\\}

We motivate this extension by considering a scenario with a primary response $Y$ exhibiting a moderate treatment effect (i.e. $U_{Y} = 0.7$) alongside two surrogate candidates, $S_{1}$ and $S_{2}$, which display strong treatment effects such that $U_{S_{1}} = 0.7$ and $U_{S_{2}} = 0.8$. Under the conventional non-inferiority testing framework, one examines whether the difference between the treatment effect on $Y$ and that on each surrogate is sufficiently less than some small positive tolerance $\epsilon$. In this setting, we obtain $\delta_{1} = U_{Y} - U_{S_{1}} = 0$ and $\delta_{2} = U_{Y} - U_{S_{2}} = -0.1$. Notably, if one were to compare these differences under the non-inferiority framework, $\delta_{2}$ would be deemed further from zero than $\delta_{1}$, potentially leading to the erroneous conclusion \---- assuming equal standard deviations (i.e. $\sigma_{\delta_{1}} = \sigma_{\delta_{2}}$) \---- that $S_{2}$ is a stronger surrogate than $S_{1}$. Intuitively, however, $S_{1}$ should be considered a better surrogate since it approximates the treatment effect on $Y$ more closely.

To overcome this inconsistency, we propose a \textit{two one-sided test procedure} \cite{Schuirmann1987} that assesses whether $\delta$ falls within the interval $\lbrack -\epsilon, \epsilon \rbrack$. For a nominal significance level $\alpha$, we define the null hypotheses as
\[
H_{0}^{(1)}: \delta \geq \epsilon \quad \text{and} \quad H_{0}^{(2)}: \delta \leq -\epsilon.
\]
The $(1-\alpha)\times 100\%$ one-sided confidence interval and the corresponding p-value for $H_{0}^{(1)}$, denoted $p^{(1)}$, have been described above. An analogous rationale applies to testing $H_{0}^{(2)}$, where the $(1-\alpha)\times 100\%$ one-sided confidence interval is given by
\[
\left[ \widehat{\delta} - \Phi^{-1}(1-\alpha)\widehat{\sigma_{\delta}},\, 1 \right],
\]
with the associated p-value defined as
\[
p^{(2)} = 1 - \Phi\!\left(\frac{\widehat{\delta} + \epsilon}{\widehat{\sigma}_{\delta}}\right).
\]
We then deem $S$ a suitable surrogate if both $H_{0}^{(1)}$ and $H_{0}^{(2)}$ can be rejected. Equivalently, this is achieved when the combined $(1-2\alpha)\times 100\%$ confidence interval,
\[
\left[ \widehat{\delta} - \Phi^{-1}(1-\alpha)\widehat{\sigma_{\delta}},\, \widehat{\delta} + \Phi^{-1}(1-\alpha)\widehat{\sigma_{\delta}} \right],
\]
is entirely contained within the interval $\lbrack -\epsilon, \epsilon \rbrack$. The overall p-value for the two one-sided test procedure is taken as
\[
p = \max(p^{(1)}, p^{(2)}).
\]

\subsection{Overview of RISE}
Our proposed approach comprises two steps. In the first, we apply the aforementioned rank-based procedure to each candidate surrogate $S_{j}$ to screen $\boldsymbol{S}$ for the most promising candidates. In the second step, we evaluate the strength of the identified set of surrogates. To avoid overfitting, we use sample splitting to separate our full data into screening and evaluation sets, such that each step uses distinct data \cite{Ball2020}.

\paragraph*{\textbf{Step 1 - Rank-Based Screening}\\}
Given a significance level $\alpha$ and desired power $(1-\beta)$, we apply the previously detailed rank-based procedure to each surrogate $S_{j}$ in the screening dataset, resulting in a point estimate $\widehat{\delta_{j}}$, its standard deviation $\widehat{\sigma}_{\delta_{j}}$, associated confidence interval and p-value. To control the excessive false discovery rate (FDR) among our identified candidate surrogates resulting from the high number of statistical tests, we perform a multiple testing correction on the p-values \cite{Bender2001}. The subset of candidate surrogates, which we call $\mathcal{S}$, can then be selected as those whose adjusted p-values fall below $\alpha$. 

\paragraph*{\textbf{Step 2 - Evaluating Strength of Surrogate}\\}
In the second step, we propose to evaluate the strength of the set $\mathcal{S}$ by first reducing the dimension of $\mathcal{S}$ to a single marker through a weighted sum
\begin{equation*}
\widehat{\gamma}_{\mathcal{S}} = \sum\limits_{j \in \mathcal{S}} \left|\widehat{\delta_{j}}\right|^{-1}\bar{S_{j}}
\end{equation*}
where $\bar{S_{j}}$ is $S_{j}$ standardised to have mean $0$ and standard deviation $1$, and the weights are the inverse of the estimated $\delta_{j}$, such that stronger surrogates contribute more to the combined marker, taking the absolute value to avoid negative weights. Then, the rank-based procedure for a single surrogate is applied with $\widehat{\gamma}_{\mathcal{S}}$ in the evaluation dataset. If the p-value falls below $\alpha$, we conclude that $\widehat{\gamma}_{\mathcal{S}}$ is a useful surrogate for $Y$. 

\subsection{Simulation Study Setup}

We conducted a simulation study to evaluate the performance of our proposed two-step procedure under varying conditions and data-generating processes. The datasets were generated with $P = 500$ variables, a nominal significance level of $\alpha = 0.05$, and results summarised over $N_{\text{sim}} = 500$ simulations. Two primary scenarios were considered, each designed to assess different aspects of performance.
In Scenario 1, no valid surrogates were generated. This setup allowed us to evaluate the false positive rate (FPR) \--- the proportion of false positives among all negatives. In Scenario 2, 10\% of the surrogates were valid, enabling the empirical assessment of the false discovery proportion (FDP)
\--- the proportion of false positives among all claimed positives \--- and the statistical power, defined as the proportion of positives found significant.
\paragraph*{\textbf{Definition of Valid Surrogates}\\}
By construction, the non-inferiority margin determines whether a variable is classified as a valid or invalid surrogate under our framework. Specifically, any $S_j$ where $U_{S_j} < U_Y - \epsilon$ is deemed invalid; otherwise, it is considered valid. The true values of $U_Y$ and $U_{S_j}$, denoted $U_Y^{*}$ and $U_{S_j}^{*}$, can be derived analytically or through the asymptotic properties of U-statistics.

For our proposed procedure, invalid surrogates were generated as perfectly useless surrogates with $\widehat{U}_{S_j} = 0.5$, and $\epsilon$ was fixed at $\widehat{U}_Y - 0.5$. This setup allowed us to examine the p-value distribution at the boundary of the non-inferiority test and investigate how increasing the strength of surrogates beyond this boundary affects the test statistical power. It should be noted that, in practice, this choice of $\epsilon$ is unlikely to be useful for identifying surrogates that explain a substantial portion of the treatment effect on $Y$.

\paragraph*{\textbf{Data-Generating Processes}\\} 
Let $p_{\text{invalid}}$ and $p_{\text{valid}}$ denote the numbers of invalid and valid surrogates, respectively. In Scenario 1, all 500 variables were invalid surrogates ($p_{\text{invalid}} = 500$ and $p_{\text{valid}} = 0$). In Scenario 2, 10\% of the variables were valid surrogates ($p_{\text{invalid}} = 450$ and $p_{\text{valid}} = 50$). We considered two different data-generating processes (DGPs).

\textbf{DGP 1: Multivariate Normal \--- }
All variables were generated from multivariate normal distributions. The responses followed $Y_a \sim \mathcal{N}(\mu_{y_a}, \sigma_{y_a})$, with $\mu_{y_1} = 3$, $\mu_{y_0} = 0$, and $\sigma_{y_a} = 1$. This setup resulted in a theoretical $U_Y^{*} = 0.985$, representing a strong treatment effect on $Y$.
Invalid surrogates were generated as $S_{j,a} \sim \mathcal{N}_{p_{\text{invalid}}}(\boldsymbol{M}, \Sigma_{\text{invalid}})$, where $\boldsymbol{M} = (m_1, \dots, m_{p_{\text{invalid}}})^T$, $m_j \sim Uniform(0.5, 2.5)$, and $\Sigma_{\text{invalid}} = \text{diag}(\sigma_1, \dots, \sigma_{p_{\text{invalid}}})$, with $\sigma_j \sim Uniform(0.5, 2)$.
Valid surrogates were generated by perturbing the true responses: $S_{j,a} = y_a + \mathcal{N}_{p{\text{valid}}}(\boldsymbol{0}, \Sigma_{\text{valid}})$, where $\Sigma_{\text{valid}} = \text{diag}(\sigma_{\text{valid}})$. The strength of surrogates was controlled by $\sigma_{\text{valid}}$, with larger values indicating weaker surrogates. Where the impact of multicollinearity was of interest, a constant $\sigma_{\text{corr}}$ was added to the off-diagonal elements of $\Sigma_{\text{invalid}}$. For valid surrogates, since the dynamic range of $\sigma_{\text{valid}}$ is large in order to explore different surrogate strengths, we take the off-diagonal elements of $\Sigma_{\text{valid}}$ to be the correlation parameter $\sigma_{\text{corr}}$ scaled by the diagonal elements i.e. $\sigma_{\text{corr}}\cdot\sigma_{\text{valid}}$.

\textbf{DGP 2: Complex Surrogate-Response Relationships \--- }
To introduce more complex invalid surrogate generation and surrogate-response relationships, responses were generated as in DGP 1, while invalid surrogates were sampled from exponential distributions: $S_{j,a} \sim \text{Exp}(\lambda_{j})$, with $\lambda_{j} \sim Uniform(0.5, 2.5)$. Valid surrogates were derived by perturbing a transformed response: $S_{j,a} = f(y_a) + \mathcal{N}_{p_{\text{valid}}}(\boldsymbol{0}, \Sigma_{\text{valid}})$, where $f(x) = x^3$, and $\Sigma_{\text{valid}}$ was as defined earlier in DGP 1.

\paragraph*{\textbf{Evaluation Stage}\\}

In the second stage of our testing procedure, some subset of markers $S_{j} \in \mathcal{S}$ are combined to form a single marker $\widehat{\gamma}_{\mathcal{S}}$. This combination may consist of both true positives and false positives, in proportions $\rho_{\text{valid}}$ and $\rho_{\text{invalid}}$, respectively. Although type I error and statistical power can be clearly defined in the case where we have either none or all false positives ($\rho_{\text{valid}} \in \{0,1\}$), it is less straightforward when the components of $\widehat{\gamma}_{\mathcal{S}}$ are mixed ($\rho_{\text{valid}} \in (0,1)$). Therefore, we opt to set $\left|\mathcal{S}\right| = 20$ and simply examine the distributions of p-values under varying $\rho_{\text{invalid}}$. Throughout, valid surrogates were generated with average strength $\widehat{U}_{S_j} = 0.9$.

\section{Results}

\subsection{Simulation Results}

\paragraph*{\textbf{Step 1 - Screening}\\}

\begin{figure*}[ht]
\centerline{\includegraphics[height=20pc]{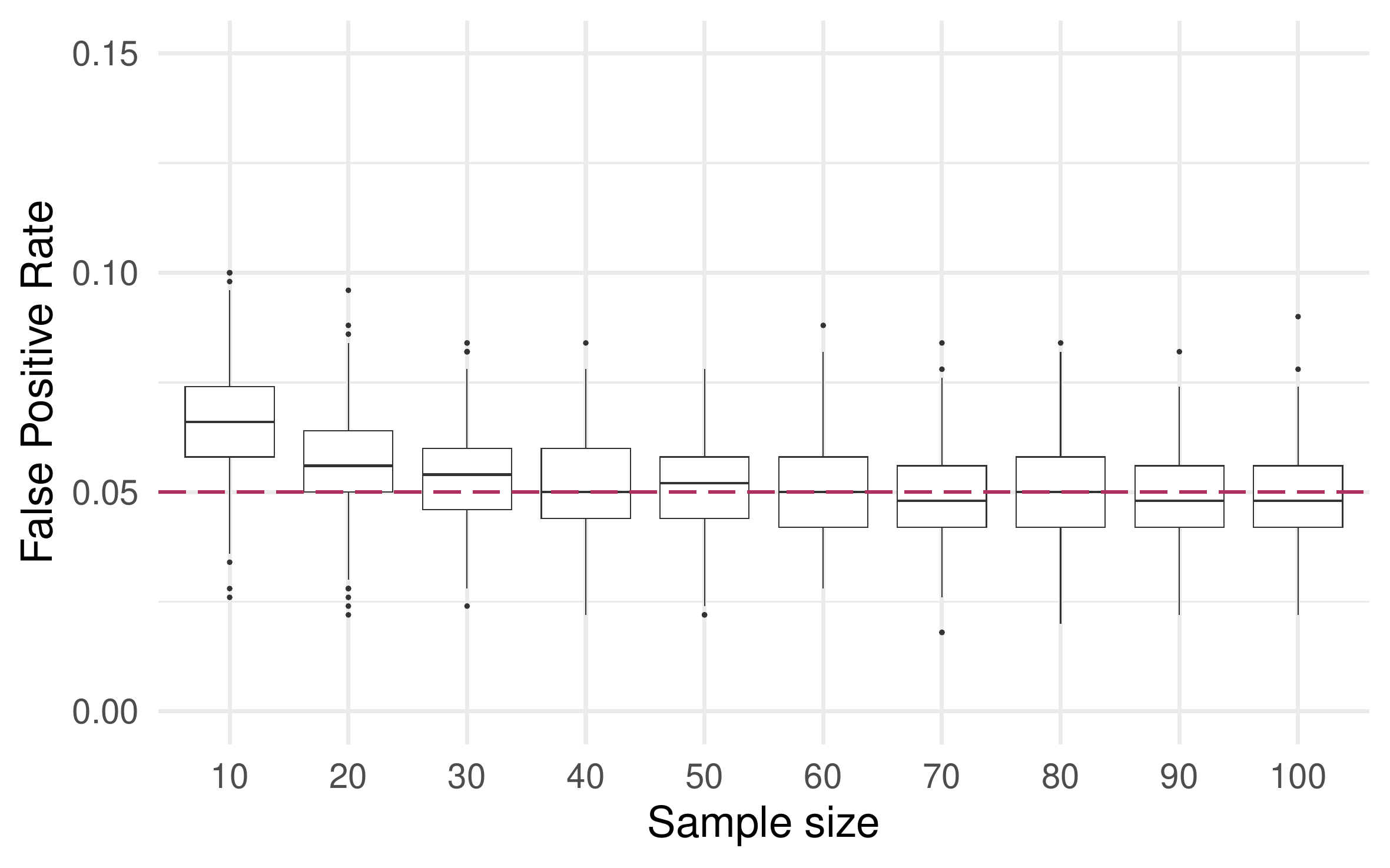}}
\caption{Data generation process 1, scenario 1: boxplots of observed false positive rates against different sample sizes in the uncorrelated setting. The nominal significance level $\alpha = 0.05$ is plotted as a dashed purple line. \label{figure1}}
\end{figure*}

\begin{figure*}[ht]
\centerline{\includegraphics[height=20pc]{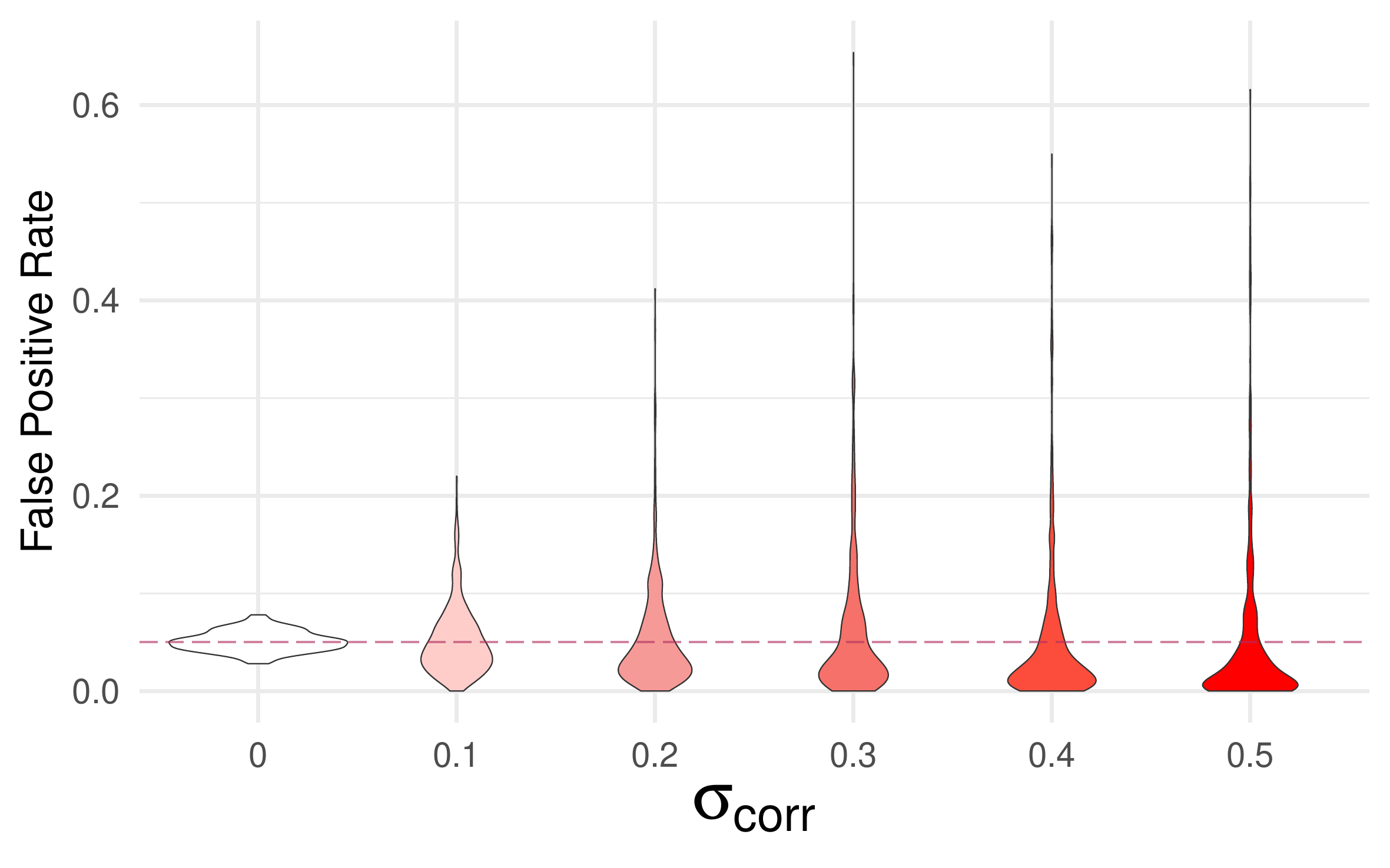}}
\caption{Data generation process 1, scenario 1: violin plots of observed false positive rates against different levels of correlation across 500 simulations for a fixed sample size of $n = 50$. Increasing the $\sigma_{\text{corr}}$ parameter increases the inter-predictor correlation. The nominal significance level $\alpha = 0.05$ is plotted as a dashed purple line. 
\label{figure2}
}
\end{figure*}

\begin{figure*}[ht]
\centerline{\includegraphics[height=20pc]{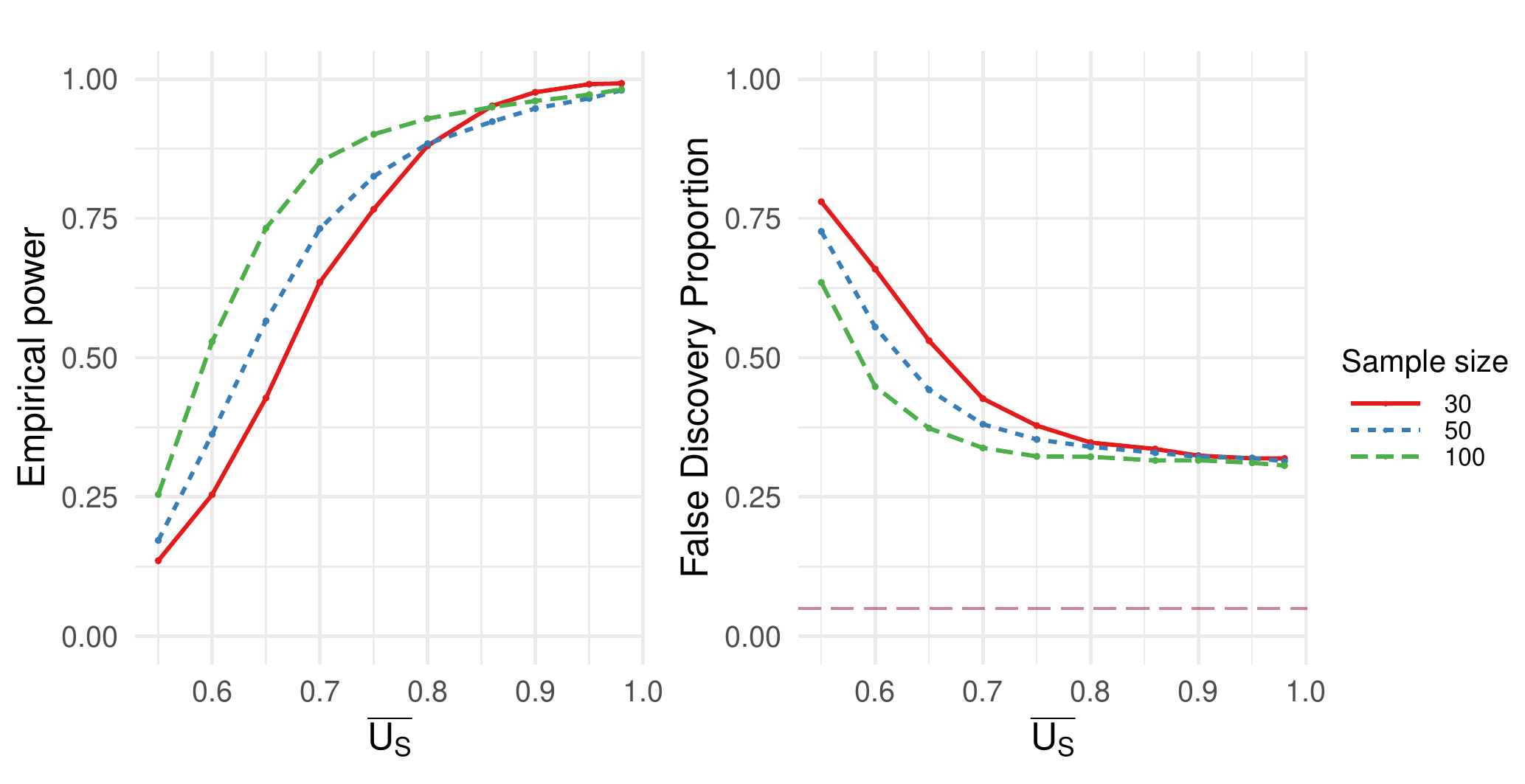}}
\caption{Data generation process 1, scenario 2: empirical power (left) and false discovery proportion (right) prior to multiple testing corrections as a function of average surrogate strength ($\bar{U_{S}}$) for three different sample sizes. The nominal significance level $\alpha = 0.05$ is plotted as a dashed purple line on the FDR plot.
\label{figure3}
}
\end{figure*}

We first examined the properties of the test under data generation process 1. In Scenario 1, where no valid surrogates were present, we examined the false positive rate (FPR) across various sample sizes in the uncorrelated setting. The empirical FPR remained close to the nominal level of 0.05 for sample sizes greater than 30, indicating a lower practical limit for the sample size and demonstrating that the procedure performs well even with small sample sizes (Figure \ref{figure1}). We then assessed the impact of correlation on the FPR for a fixed sample size of $n = 50$. In the absence of correlation, the FPR remained close to the nominal value with minimal variance across simulations. However, as correlation increased, the mean FPR decreased below the nominal value of 5\% but its variance grew, with the highest correlation levels leading to a handful of extreme outliers (Figure \ref{figure2}). In Scenario 2, where there was 10\% of valid surrogates, we evaluated the empirical FDP and empirical power (or true positive rate) across varying surrogate strength values ($\widehat{U}_{S} = 0.55, 0.60, \dots, 0.95, U_Y$) and sample sizes ($n = 30, 50, 100$) in the uncorrelated setting. As expected, empirical power increased to nearly 1, while empirical FDR decreased to its minimum value as the average surrogate strength increased (Figure \ref{figure3}). When examining the impact of correlation in Scenario 2 for a fixed sample size of $n = 50$ and average surrogate strength of $\widehat{U}_{S} = 0.9$, we found that the FDP decreased on average, but became more variable at higher correlation levels. However, empirical power appeared to be largely unaffected by correlation (Supporting Information Figure S1). We also assessed the effect of three multiple testing correction methods: Benjamini-Hochberg (B-H), Bonferroni, and Benjamini-Yekutieli (B-Y) \cite{Benjamini1995, Dunn1961, Benjamini2001}. As expected, all three of the procedures controlled the FDR well and resulted in satisfactory power at high surrogate strengths (Supporting Information Figure S2). The Bonferroni and B-Y procedures were found to offer stricter control of the FDR compared to the B-H procedure, which provided more balance between controlling the expected FDR and maintaining the power to detect true signals.

We next examined the properties of the test under the more complex data generation process 2. Overall, the properties of the test remained similar to those observed under DGP 1, with the only notable difference being in the observed FPR, which was more stable across different levels of inter-predictor correlation (Supporting Information Figures S3, S4, S5).

\paragraph*{\textbf{Step 2 - Evaluation}\\}

\begin{figure*}[ht]
\centerline{\includegraphics[height=20pc]{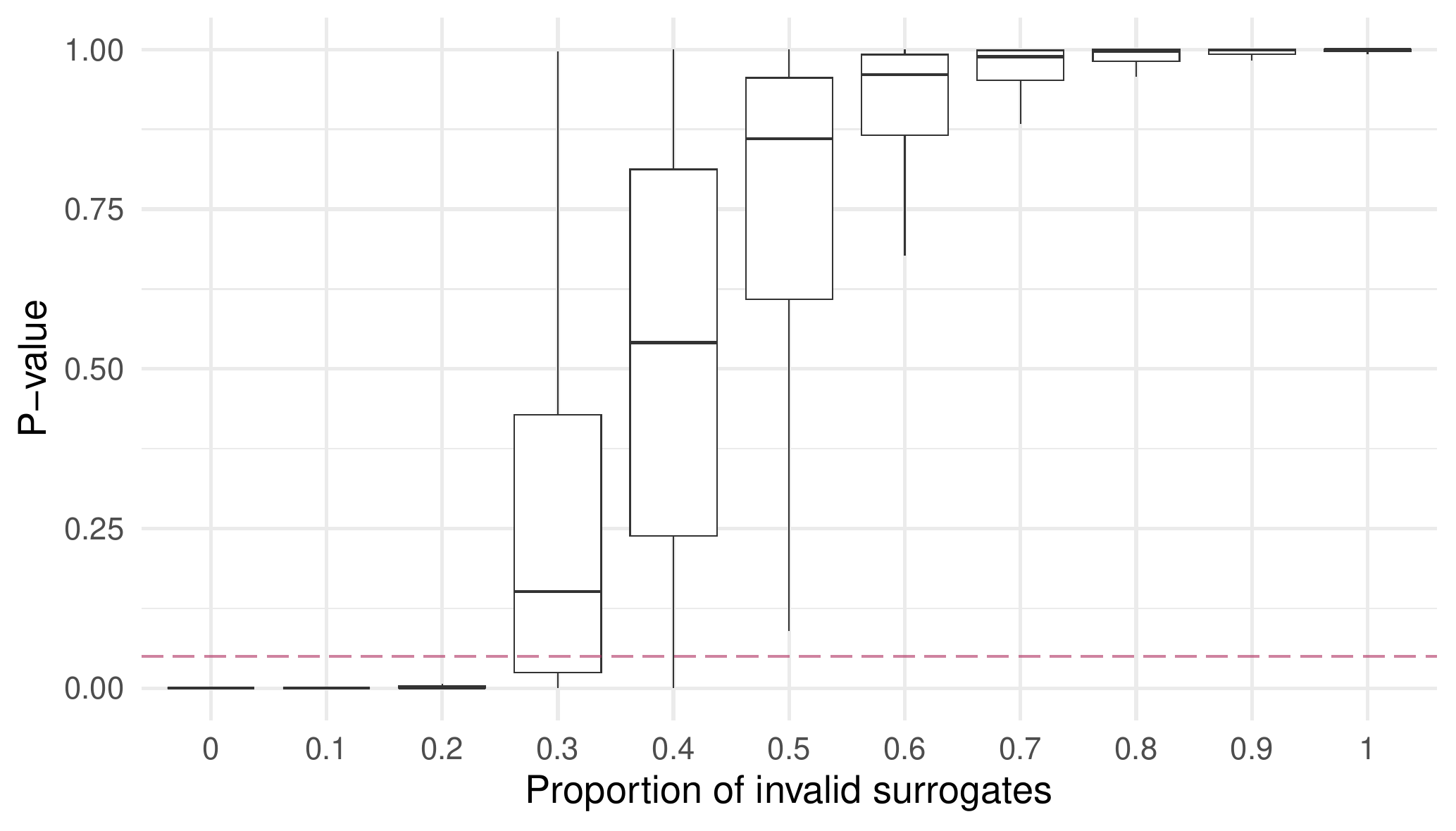}}
\caption{Data generation process 1: the distributions of the p-values in the evaluation step are examined as a function of the false discovery proportion which make up $\widehat{\gamma}_{\mathcal{S}}$, which consists of a combination of $20$ predictors. The sample size is $n = 50$ and the valid surrogate strength is $\widehat{U_{S_{j}}} = 0.9$. The nominal significance level $\alpha = 0.05$ is plotted as a dashed purple line. Desired power for the new surrogate was fixed at $80\%$. 
\label{figure4}
}
\end{figure*}

In the evaluation stage, we examined the distribution of p-values as a function of the FDP in $\widehat{\gamma}_{\mathcal{S}}$ for a fixed sample size $n = 50$. For both data generation processes, When the FDP was low ($\leq 0.2$), the null hypothesis was always rejected (indicating that $\widehat{\gamma}_{\mathcal{S}}$ was a strong surrogate). In contrast, the null hypothesis was never rejected when the FDP was too high ($\geq 0.6$). When $\widehat{\gamma}_{\mathcal{S}}$ contained a balanced mixture of false and true positives ($0.3 < \rho_{\text{invalid}} \leq 0.5$), the null hypothesis was mostly not rejected, but p-values exhibited higher variance (Figure \ref{figure4}, Supporting Information Figure S6). This is desirable behaviour, as we have shown the false discovery proportion to be low in our setup subject to the 3 multiple testing corrections tested (Supporting Information Figure S2), which will lead to rejection of the null hypothesis for $\widehat{\gamma}_{\mathcal{S}}$. In addition, in the event of an elevated false discovery proportion, the null hypothesis is unlikely to be rejected. These conclusions were also found to hold under smaller and larger sample sizes (Supporting Information Figure 9).

\subsection{Application to Influenza Vaccination Data}

We applied RISE to publicly available gene expression and immune response data to identify and evaluate potential surrogate markers of the immune response to the trivalent inactivated influenza vaccine (TIV). These data are available from the ImmPort platform \cite{Bhattacharya2018} (immport.org) under study accession number SDY1276 (entited \textit{time series of global gene expression after trivalent influenza vaccination in humans}). TIV is a seasonal flu vaccine containing inactivated forms of three influenza virus strains, designed to stimulate immune protection without causing infection \cite{Wong2013}. We applied RISE to a study examining the response of young, healthy adult volunteers to the 2008-2009 TIV formulation, which is designed to protect against two strains of influenza A and one strain of influenza B. Due to the previously reported variability in response to influenza vaccination at both the immune and transcriptomic levels based on sex \cite{Furman2013, Wen2018}, we further subsetted the data to include only female subjects. 

Since this vaccine targets three strains of influenza, neutralising antibody response data were available for each strain. To obtain a single measure of neutralising antibody titres, we computed the mean across the three strains for each individual. An alternative could have been to study titres for each strain separately \cite{Kandinov2025}. Surrogate candidates were defined as gene expression levels in whole blood cells, as assessed using microarray technology.

Although these data lack a placebo control, measurements were collected on the same individuals both before and after vaccination. This is therefore a paired data setting where each individual provides response and surrogate candidate values, namely
\[
\boldsymbol{Y}_i = \left(Y_{i}^{\text{pre-vaccine}},\, Y_{i}^{\text{post-vaccine}}\right)^{T} \qquad \text{and} \qquad \boldsymbol{S}_i = \left(S_{i}^{\text{pre-vaccine}},\, S_{i}^{\text{post-vaccine}}\right)^{T}.
\]

Pre-vaccination measurements were taken on day 0 (i.e. immediately prior to vaccination), while post-vaccination measurements were taken on day 28 for the response and on day 1 for the surrogate candidates. The objective was to determine whether the average vaccine effect on any gene expression markers, observed one day post-vaccination, could predict the average vaccine effect on the neutralising antibody titres, observed 28 days post-vaccination. 

The dataset comprised paired observations on $n = 103$ individuals.  Due to the fact that the screening stage involves a large number of tests, necessitating adjustment of the resulting p-values to control the false discovery rate, we commit the majority of our data to the screening phase, splitting our data randomly into screening and evaluation datasets at a ratio of $75$:$25$, respectively.

The significance level was chosen as $\alpha = 0.05$, and the desired power in the screening stage was fixed at $90\%$. The Bonferroni procedure was used to correct the resulting p-values. The paired data extension as well as the two one-sided test procedure were used (Section 2.3). The estimated value of $U_Y$ was $0.97$, reflecting a strong neutralising antibody response to TIV, and corresponding to a non-inferiority threshold of $\epsilon = 0.29$. Among the $10,086$ genes in the expression data, 222 had an adjusted p-value below $0.05$. For brevity, we display the genes with the smallest adjusted p-values in Table \ref{table1}, along with the estimates for $U_Y - U_{S_j} = \delta_j$, corresponding two-sided $90\%$ confidence intervals, standard deviations $\widehat{\sigma_{\delta_j}}$, and both raw and adjusted p-values. The full table of genes can be found in Supporting Information Table S1.

\begin{figure*}[ht]
\centerline{\includegraphics[height=20pc]{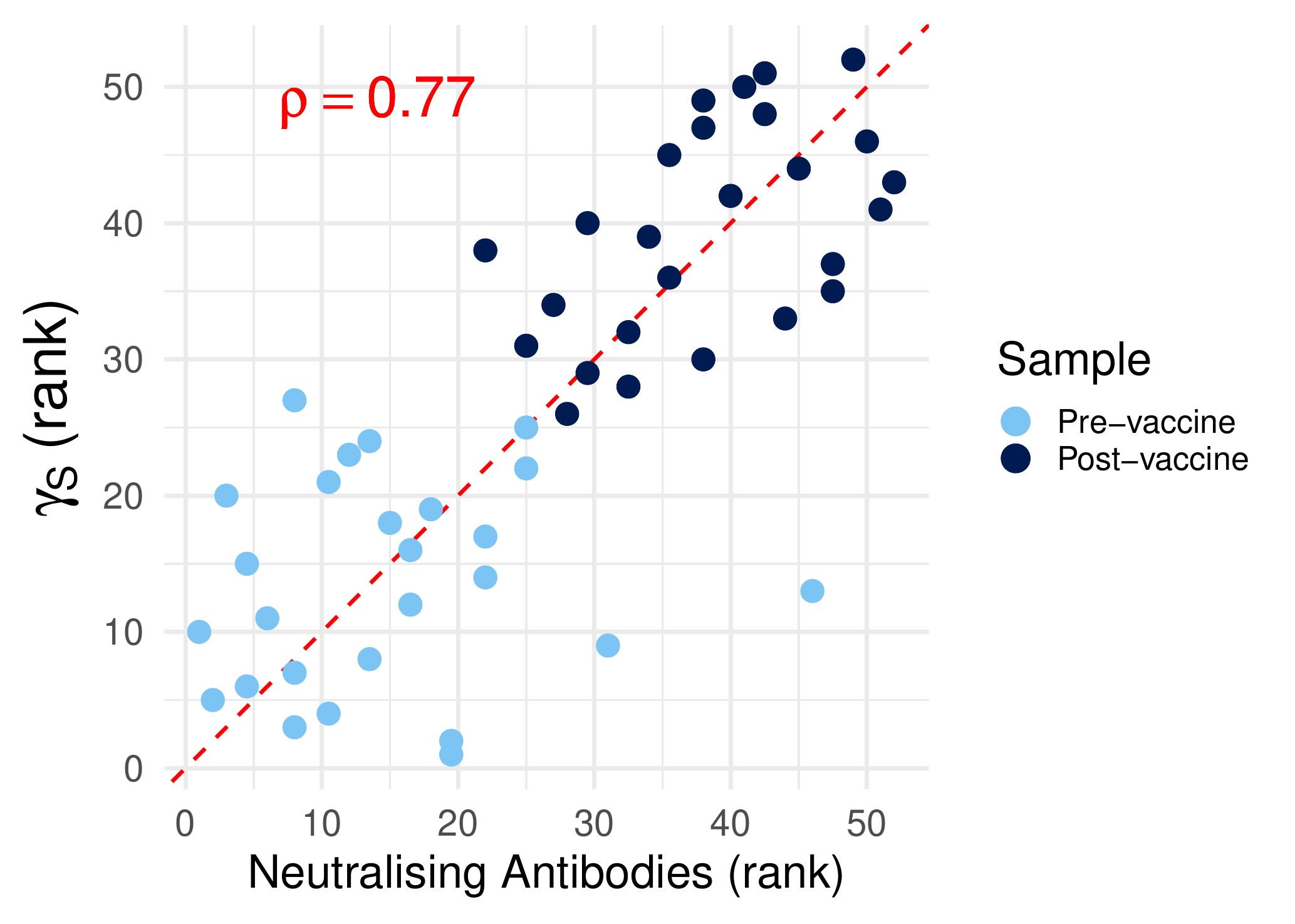}}
\caption{Ranks of the mean cross-strain neutralising antibodies against the ranks of the constructed 222-gene-combination surrogate marker in the evaluation dataset. The Spearman rank correlation coefficient is 0.77, indicating strong positive correlation.
\label{figure5}
}
\end{figure*}

In the evaluation phase, the identified set of $222$ genes from Step 1 of RISE were combined using a standardised weighted sum to form a single predictor, denoted $\gamma_S$. In the evaluation dataset, the estimated value of $U_Y$ was $0.96$, corresponding to $\epsilon  = 0.14$ for a desired $0.90$ powered test based on $\gamma_S$. The value of $\delta$ was found to be $-0.038$ ($95\%$ C.I. $[-0.10, 0.025]$), yielding a p-value of $0.003$. The negative value of $\delta$ reflects the fact that the point estimate of the treatment effect on the combined surrogate is slightly stronger than that on the antibody response. These results suggest that the constructed $\gamma_S$ is a reasonable trial-level surrogate for the neutralising antibody response to the 2008-2009 TIV amongst females. This is further illustrated in Figure \ref{figure5}, which plots the ranks of the true response against $\gamma_S$, showing strong positive correlation between the antibody ranks and the new surrogate ranks (Spearman rank correlation coefficient $\rho= 0.77$).

In our application, the non-inferiority threshold for screening was large, even at a desired power of 0.90, resulting in a high number of significant genes. This outcome is unsurprising given the large effect size of the primary outcome and the larger sample size. To control for multiple testing, we applied the Bonferroni correction, the strictest method considered, which consequently yielded a more parsimonious gene signature. We additionally conducted a sensitivity analysis to evaluate the robustness of the results to the value of $\epsilon$ in the screening stage (Supporting Information Table S2). Choosing stricter tolerance levels resulted in more parsimonius signatures, however, the strength metric $\delta_{\gamma_{\mathcal{S}}}$ of the combination of these signatures on the evaluation data did not vary. In addition, we compared the evaluation metrics between $\delta_{\gamma_{\mathcal{S}}}$ and the top genes from the screening phase individually (Supporting Information Table S3). Each of the top 10 genes resulted in evaluation metrics similar to that of the combined marker. These results likely reflect the high degree of redundant information between genes in similar biological pathways as well as the fact that the top genes may dominate the surrogacy of the combined surrogate. Indeed, the biological functions of the 222 genes were examined using DAVID bioinformatics to identify ontological terms which were significantly over-represented in the list. This revealed a significant proportion of these genes to be related to innate antiviral processes, providing a clear immunological interpretation of the signature (Table \ref{table2}); we discuss this further in the Discussion. 

In conclusion, we identified a subset of genes whose early post-vaccination expression may serve as a promising surrogate for the mid-term immunogenicity of an inactivated influenza vaccine in healthy adult females. This provides a basis for further validation and illustrates RISE’s practicality as a framework for exploring trial-level surrogate markers in clinical studies with high-dimensional candidate markers.

\section{Discussion}

Surrogate markers can provide significant advantages in the conduct of randomised clinical trials, particularly those evaluating vaccine immunogenicity. High-dimensional molecular markers are promising candidates for surrogates in this context due to their biological relevance and practical utility. However, existing methods for identifying and validating surrogate markers typically break down in high-dimensional contexts, necessitate large sample sizes, or rely on restrictive parametric assumptions. In this study, we introduced RISE \--- a novel two-step method for identifying and evaluating high-dimensional surrogate markers, applied in the context of a vaccine clinical trial.

Our approach builds upon existing rank-based methodologies by adapting them for high-dimensional settings through a combination of univariate testing and dimension reduction, followed by evaluation using independent data. RISE effectively addresses several key challenges associated with evaluating high-dimensional molecular surrogates, such as the large number of candidate surrogates, limited sample sizes, and the need for false discovery rate control. The initial screening step utilises a non-parametric, rank-based univariate test to evaluate whether each variable approximates the treatment effect on the response within some small margin. As discussed in the study introducing the rank-based approach for single surrogate markers \cite{Parast2024}, this method offers several advantages that are particularly relevant in high-dimensional vaccine trials. First, the test enables robust and valid inference even in small sample scenarios, where assumptions like linearity and normality are difficult to verify. Second, being rank-based, the test is invariant to data transformations and robust to outliers. Finally, by comparing entire rank distributions rather than relying on summary statistics like the mean, this method provides a more comprehensive assessment of surrogate strength. The evaluation step in RISE then uses a weighted combination of the screened predictors to form a synthetic biomarker. 

We underscore that the primary objective of RISE is to evaluate trial-level surrogacy, which concerns the prediction of average treatment effects. While we applied the sample‐splitting strategy to assess the internal validity of our derived surrogate, the approach is readily generalisable for evaluating cross‐trial validity when multi‐trial data are available. For example, promising surrogate candidates may be identified using data from one trial and then validated individually in other trials; a surrogate is deemed to generalise well if the relationship between the estimated treatment effects on Y and the derived surrogate can be reliably predicted across trials. We see RISE as a tool which could be used as one part of a comprehensive strategy for assessing surrogacy. For instance, RISE could be applied to high-dimensional data to derive a univariate marker, which could then be further explored with low-dimensional surrogacy methods using alternative frameworks such as principal stratification \cite{Gilbert2008}, causal inference \cite{Laan2011}, or meta-analysis \cite{Molenberghs_2002}.

Although RISE was developed for continuous outcomes and surrogate markers, it could be applied to various data types with some modifications. The spirit of the approach is in the comparison of non-parametrically estimated U-statistics between a surrogate candidate and an outcome. This requires the data to have some kind of natural ordering of values, allowing some to be considered ``better" than others. RISE could therefore be extended easily to ordinal (e.g.\ antibody levels from semi-quantitative assays) and binary outcomes (e.g.\ infected vs.\ uninfected after a given period)--- either with adjustments to the current framework to account for ties, or by targeting a different measure of association (e.g. Kendall's Tau). Equally, a natural extension to time-to-event data would be to use log-rank statistics as the measure of treatment effect on survival outcomes and/or surrogates.

As with any epidemiological analysis, it is important to consider aspects such as confounding and missing data when applying RISE. For example, there are often baseline covariates such as age and sex which may confound the relationships between the surrogates and the outcome. We clarify that, in the marginal randomised controlled trial setting, these variables may not be considered surrogates in themselves, as they are, by design, not associated with treatment. Nevertheless, it may be still be of interest to consider the implications of such pre-treatment confounders as well as potential heterogeneity in surrogate strength with respect to baseline surrogates, as investigated in recent works \cite{Parast2021, Knowlton2025}. If there are residual imbalances in potentially confounding baseline factors post-randomisation, conventional adjustment methods—such as restriction, regression modelling or inverse‑probability weighting—can be applied to control for these effects. The handling of missing data depends on the supposed mechanism of missingness. If data are missing completely at random (MCAR), complete-case analysis may be appropriate without introducing bias. For data missing at random (MAR), multiple imputation using observed covariates can help preserve comparability between the two U-statistics. In cases of data missing not at random (MNAR), more advanced modelling approaches are required, and caution is warranted when interpreting results. 

Our simulation studies illustrate the favourable properties of RISE’s screening and evaluation procedures. We demonstrated that the test procedure is valid and well-calibrated, although caution is required for very small samples or when inter-predictor correlations are high, as the false positive rate may slightly deviate from the nominal level. Additionally, we explored how surrogate strength and sample size influence empirical power and false positive rates. The test exhibited high power to detect true positives and minimised false positives when surrogates were strong, even with small sample sizes. This is encouraging, as in practice, we are primarily concerned with identifying the strongest surrogates. These findings also emphasise the importance of multiple testing corrections in order to control the elevated false positive rate in situations with a low proportion of true positives amongst high-dimensional predictors. We demonstrated these results under simple Gaussian data generation as well as in a more complex setting. While these scenarios were chosen to capture a range of plausible conditions, we acknowledge that they constitute only a limited subset of potential surrogate–response relationships. Based on these results, we envisage the approximate range of sample size for the application of RISE to be $n = 30-200$. Although there is nothing which stops one applying RISE in larger sample sizes, other methods may exist which may be better powered or provide stronger forms of evidence for surrogate validity \cite{Parast_2015}.

In applying RISE to a vaccine trial, our objective was to identify early gene expression markers that could serve as surrogates for the neutralising antibody response following immunisation with a seasonal trivalent inactivated influenza vaccine (TIV). A signature of 222 genes was identified, whose expression appeared to function individually as effective surrogates in the screening data subset. A standardised, weighted combination of these 222 genes was then evaluated on independent data as a viable candidate to replace the day 28 neutralising antibody response. The biological functions of these genes were explored using DAVID \cite{Huang2008}, a bioinformatics tool that summarises biological functions associated with a gene group by identifying over-represented terms compared to those expected by random sampling of the same number of genes. This analysis revealed that many genes in the signature were linked to antiviral defence and innate immunity pathways (Table \ref{table2}). In particular, numerous genes in the list are known to regulate or be stimulated by interferons, a family of proteins that interfere with viral infections, making these genes sensitive indicators of innate immune activation. Genes in this signature related to the interferon response include interferon-induced genes (IFI16, IFI35, IFI44, IFI44L, IFI6, IFIH1, IFIT1, IFIT2, IFIT3, IFITM1, IFITM3), the STAT family (STAT1, STAT2, STAT5A), interferon regulatory factors (IRF1, IRF2, IRF5, IRF7, IRF9), the OAS family (OAS1, OAS2, OAS3, OASL), MX proteins (MX1, MX2), viral sensors (DDX58, DDX60), and other interferon-stimulated genes (ISG20, GBP1, GBP2, BST2, RSAD2, XAF1, TRIM21, TRIM22, TRIM5). Further work is required to determine whether activation of these pathways can robustly predict the neutralising antibody response at both the trial and individual levels, across individuals with varying instrinsic characteristics and different formulations of TIV.

While this study represents a significant advancement in non-parametric methods for identifying high-dimensional surrogate markers, several limitations must be acknowledged. One such limitation is the criteria we propose for evaluating trial-level surrogacy. We define our treatment effects $U_{Y}, U_{S}$ on the probability scale and estimate these using non-parametric rank-sum statistics. While this offers many advantages which are highlighted above, a notable disadvantage is that it does not distinguish between two surrogate markers which both separate treatment groups perfectly but which have differing magnitudes of effect on the natural scale (as both would have $U_{S} = 1$). The other difficulty associated with our criteria is in the selection and interpretation of the non-inferiority margin $\epsilon$. While we propose a data-driven approach linking $\epsilon$ to the sample size and desired power, it remains ad-hoc and challenging to interpret. Our recommendation is that users perform sensitivity analyses to evaluate the study's conclusions on the value of $\epsilon$.  

Another methodological limitation is the manner in which we propose combining candidate markers into a single surrogate. We propose a weighted, standardised sum of the markers that pass the screening stage, with weights proportional to their strength as surrogates. While this approach is intuitive, one may argue that the resulting surrogate has limited biological interpretation and lacks certain optimality conditions as proposed by others in the literature \cite{Wang2022a}.

The choice of outcome in our data application was the neutralising antibody levels, a continuous marker whose quantity and duration has been used itself as a surrogate marker to study the efficacy of TIV. This results in the identification of markers which are, in reality, surrogates of a surrogate. While this approach is pragmatic in the absence of true clinical endpoints, it introduces limitations, including a potentially weaker or indirect relationship with the ultimate outcome of interest, and the risk of identifying markers that lack generalisability or biological relevance to true disease protection. It is also crucial to emphasise that the markers identified by RISE are statistical surrogates rather than mechanistic ones, meaning they are associated with both the vaccine and its induced neutralising antibody response, but may not necessarily directly reflect the underlying biological mechanisms \cite{Plotkin2012}. This distinction is vital in the context of gene expression studies, where complex co-expression patterns and regulation cycles may lead to variables associated with an intervention and its outcome, but which are not directly involved in the mechanism. 

Finally, a limitation of the univariate screening stage in RISE is that it may overlook more complex multivariate surrogates. In the gene expression context, it is well recognised that genes operate within biological pathways rather than in isolation \cite{PitaJuarez2018}. Consequently, even if no single gene qualifies as a candidate surrogate via univariate screening, a combination of genes might collectively serve as a strong surrogate. One way to address this issue is to redefine the units of investigation as groups of biologically related genes. These "gene sets" can be defined either based on prior biological knowledge or through data-driven approaches that identify co-expressed genes under specific biological conditions \cite{Altman2021, Li2013}. New surrogate candidates can then be constructed by summarising the expression levels of the genes within each set using an appropriate summary statistic (e.g., the mean, interquartile range, or maximum of standardised expression values). Subsequently, RISE can be applied to these aggregated variables, thereby increasing the likelihood of identifying biological pathways that serve as effective surrogates even when none of the individual genes do so on their own. In addition, gene expression responses may demonstrate large differences in temporal dynamics between individuals. Directions for future development to improve the RISE methodology therefore include extension to the multivariate setting to account for pathway-level trends, as well as the consideration of more complex experimental designs, such as longitudinal measurements of high-dimensional surrogate candidates.

\section*{Supporting Information}

Full supporting information may be found below. R markdown files to fully reproduce all results from this article are available on the GitHub repository \textit{github.com/arthurhughes27/RISE-project}. Functions and documentation to apply RISE are available in the R package \textit{SurrogateRank}, available on the CRAN.

\section*{Acknowledgements}

This work is part of A.H.'s PhD thesis at the University of Bordeaux, co-supervised by B.P.H. and R.T., and supported by the University of Bordeaux's Digital Public Health Graduate School, funded by France's PIA 3 scheme (Investments for the Future – Project reference: 17-EURE-0019) through the Agence Nationale de la Recherche as well as University of Bordeaux’s France 2030 program / RRI PHDS. This work was supported by the Programme et Equipement Prioritaire de Recherche Santé Numérique (PEPR SN) project SMATCH (ref: 22-PESN-0003), by INRIA's DESTRIER associate team from the  Inria@SiliconValley program, and by the NIDDK grant R01DK118354. The authors thank the Human Immunology Project Consortium for providing the ImmPort platform for open-access immunology data. We also acknowledge the investigators who contributed data for study SDY1276, supported by the NIAID Program Research Project Grant (Parent R01) and the Viral Respiratory Pathogens Research Unit (VRPRU)-266030039. The authors benefited from the Rank-Based Identification of Surrogates in Small Ebola Studies (RISE) Symposium, organized by L.P. and B.P.H. at the University of Texas at Austin, USA. This event was supported by the Dr. Cecile DeWitt-Morette France-UT Endowed Excellence Fund, which promotes scholarly collaborations between UT Austin and institutions in France. The authors thank Laura Richert for her feedback about the handling of paired samples. The authors disclose that generative AI tools were used to refine the phrasing of certain sections of this manuscript. The authors have verified that the usage of these tools did not impact the content, nor its accuracy, and has solely been used to render the article more readable. 

\bibliographystyle{unsrt}  


\newpage

\begin{table*}[ht]%
\centering %
\caption{Screening results from the data application - top genes by adjusted p-values.}
\label{table1}%
\begin{tabular*}{\textwidth}{@{\extracolsep\fill}lllll@{\extracolsep\fill}}
\toprule
\textbf{Gene} & $\boldsymbol{\delta}$ \textbf{(95\% C.I.)} & $\boldsymbol{\sigma_{\delta}}$ & \textbf{Unadjusted p-value} & \textbf{Bonferroni Adjusted p-value} \\
\midrule
  CNDP2 & -0.026 (-0.056, 0.004) & 0.018 & 1.6e-47 & 1.6e-43 \\ 
  IFI44L & -0.026 (-0.056, 0.004) & 0.018 & 1.6e-47 & 1.6e-43 \\ 
  IFITM3 & -0.026 (-0.056, 0.004) & 0.018 & 1.6e-47 & 1.6e-43 \\ 
  NPC2 & -0.026 (-0.056, 0.004) & 0.018 & 1.6e-47 & 1.6e-43 \\ 
  PSME1 & -0.026 (-0.056, 0.004) & 0.018 & 1.6e-47 & 1.6e-43 \\ 
  SERPING1 & -0.026 (-0.056, 0.004) & 0.018 & 1.6e-47 & 1.6e-43 \\ 
  VAMP5 & -0.026 (-0.056, 0.004) & 0.018 & 1.6e-47 & 1.6e-43 \\ 
  EPB41L3 & -0.013 (-0.05, 0.024) & 0.023 & 1.1e-34 & 1.1e-30 \\ 
  IFI6 & -0.013 (-0.05, 0.024) & 0.023 & 1.1e-34 & 1.1e-30 \\ 
  IRF7 & -0.013 (-0.05, 0.024) & 0.023 & 1.1e-34 & 1.1e-30 \\ 
  MX1 & -0.013 (-0.05, 0.024) & 0.023 & 1.1e-34 & 1.1e-30 \\ 
  MYOF & -0.013 (-0.05, 0.024) & 0.023 & 1.1e-34 & 1.1e-30 \\ 
  OAS3 & -0.013 (-0.05, 0.024) & 0.023 & 1.1e-34 & 1.1e-30 \\ 
  PSMB9 & -0.013 (-0.05, 0.024) & 0.023 & 1.1e-34 & 1.1e-30 \\ 
  RHBDF2 & -0.013 (-0.05, 0.024) & 0.023 & 1.1e-34 & 1.1e-30 \\ 
  SCO2 & -0.013 (-0.05, 0.024) & 0.023 & 1.1e-34 & 1.1e-30 \\ 
  UBE2L6 & -0.013 (-0.05, 0.024) & 0.023 & 1.1e-34 & 1.1e-30 \\ 
  WARS1 & -0.013 (-0.05, 0.024) & 0.023 & 1.1e-34 & 1.1e-30 \\ 
\bottomrule
\end{tabular*}
\end{table*}

\begin{table*}[ht]%
\centering %
\caption{Functional annotation analysis of 222 significant genes using DAVID. The table lists the ten most enriched functional terms, along with the number of genes associated with each term and their corresponding adjusted p-values. Each adjusted p-value reflects the significance of term enrichment assuming the null hypothesis of a random sampling of the same number of genes from the genes under study.}
\label{table2}%
\begin{tabular*}{\textwidth}{@{\extracolsep\fill}lll@{\extracolsep\fill}}
\toprule
\textbf{Term} & \textbf{Number of genes} & \textbf{B-H Adjusted p-value} \\
\midrule
  KW-0051-Antiviral defense & 36 & 5.8e-33 \\ 
  GO:0051607-defense response to virus & 41 & 1.7e-29 \\ 
  KW-0391-Immunity & 59 & 1.0e-29 \\ 
  KW-0399-Innate immunity & 44 & 1.5e-22 \\ 
  GO:0009615-response to virus & 26 & 3.9e-18 \\ 
  GO:0045087-innate immune response & 41 & 3.4e-17 \\ 
  GO:0045071-negative regulation of viral genome replication & 18 & 7.9e-17 \\ 
  GO:0140374-antiviral innate immune response & 13 & 1.3e-08 \\ 
  hsa05164:Influenza A & 20 & 2.6e-08 \\ 
  GO:0034341-response to type II interferon & 10 & 3.6e-08 \\
\bottomrule
\end{tabular*}
\end{table*}

\clearpage
\begin{center}
\textbf{\Large Supporting Information\\[0.5em] RISE: Two-Stage Rank-Based Identification of High-Dimensional Surrogate Markers Applied to Vaccinology}
\end{center}

\setcounter{equation}{0}
\setcounter{figure}{0}
\setcounter{table}{0}
\setcounter{section}{0}
\makeatletter
\renewcommand{\theequation}{S\arabic{equation}}
\renewcommand{\thefigure}{S\arabic{figure}}
\renewcommand{\thetable}{S\arabic{table}}
\renewcommand{\thesection}{S\arabic{section}}
\makeatother

\section{Variance derivations for the paired-sample extension}

In this setting, observed data consist of $i = 1,...,n$ i.i.d observations of primary response $\boldsymbol{Y_{i}} = (Y_{i}^{1},Y_{i}^{0})^{T}$ and surrogate candidate $\boldsymbol{S_{i}} = (S_{i}^{1},S_{i}^{0})^{T}$. As before, define treatment effects on $Y$ and $S$ respectively as 

$$U_{Y} = \mathbb{P}(Y^{1} > Y^{0}) + \frac{1}{2}\mathbb{P}(Y^{1} = Y^{0})$$

$$U_{S} = \mathbb{P}(S^{1} > S^{0}) + \frac{1}{2}\mathbb{P}(S^{1} = S^{0}),$$

Since data are paired, we estimate these by the proportion of individuals for which the treated observation is greater than its control counterpart, i.e. 

$$\widehat{U}_{Y} = n^{-1}\mathlarger{\sum}\limits_{i=1}^{n}G(Y_{i}^{1}, Y_{i}^{0})\qquad$$

$$\widehat{U}_{S} = n^{-1}\mathlarger{\sum}\limits_{i=1}^{n}G(S_{i}^{1}, S_{i}^{0})\qquad$$

where 

$$G(A,B) = \begin{cases}
  1 , \quad \mbox{if}\quad  A> B\\
  \frac{1}{2}, \quad \mbox{if} \quad A=B\\
  0, \quad \mbox{if}\quad  B<A
\end{cases}$$

We first derive the variance of the estimated U-statistics $\widehat{U}_{Y}$ (and by extension $\widehat{U}_{S}$) under the null hypothesis of no treatment effect.  

First, note that we only have three possible events : $Y^{1} > Y^{0}$, $Y^{0} > Y^{1}$, $Y^{0} = Y^{1}$. Under the null hypothesis, $P(Y^{1} > Y^{0}) = P(Y^{1} < Y^{0})$. Let $P(Y^{0} = Y^{1}) = \pi$. Now, since the sum of the probability of all three events must be 1, we have

$$\pi = 1- 2P(Y^{1} > Y^{0})$$ 

$$\implies P(Y^{1} > Y^{0}) = \ddfrac{1-\pi}{2}$$

Now, 
\begin{align*}
  E(G(Y^{1}, Y^{0})) &=  1\cdot P(Y^{1} > Y^{0}) + \frac{1}{2}\cdot P(Y^{1} = Y^{0}) + 0 \cdot P(Y^{1} < Y^{0}) \\
                     &= \ddfrac{1-\pi}{2} + \frac{\pi}{2} \\
                     &= \frac{1}{2}
\end{align*}

To derive the second moment, notice that 

$$G(A,B)^{2} = \begin{cases}
  1 , \quad \mbox{if}\quad  A> B\\
  \frac{1}{4}, \quad \mbox{if} \quad A=B\\
  0, \quad \mbox{if}\quad  B<A
\end{cases}$$

Then, 

\begin{align*}
  E(G(Y^{1}, Y^{0})^{2}) &=  1\cdot P(Y^{1} > Y^{0}) + \frac{1}{4}\cdot P(Y^{1} = Y^{0}) + 0 \cdot P(Y^{1} < Y^{0}) \\
                     &= \ddfrac{1-\pi}{2} + \frac{\pi}{4} \\
                     &= \ddfrac{2-\pi}{4}
\end{align*}

So, the variance is

\begin{align*}
  Var(G(Y^{1}, Y^{0})) &= E(G(Y^{1}, Y^{0})^{2}) - E(G(Y^{1}, Y^{0}))^{2} \\
                       &= \ddfrac{2-\pi}{4} - \frac{1}{4} \\ 
                       &= \ddfrac{1-\pi}{4}
\end{align*}

Then, since individuals are independent, we have 

\begin{align*}
  Var(U_{Y}) &= Var(\ddfrac{1}{n} \mathlarger{\sum\limits_{i = 1}^{n}}G(Y_{i}^{1}, Y_{i}^{0})) \\
             &= \ddfrac{n}{n^{2}}  Var(G(Y^{1}, Y^{0})) \\ 
             &= \ddfrac{1-\pi}{4n}
\end{align*}

In the case of a truly continuous response, we have $\pi = 0$ and $Var(U_{Y}) = \ddfrac{1}{4n}$. Otherwise, in the case of ordinal responses we can estimate 

$$\widehat{\pi} = \ddfrac{1}{n}\mathlarger{\sum\limits_{i =1}^{n}} \mathbbm{1}(Y_{i}^{1} = Y_{i}^{0})$$

The estimated null variance of the U-statistics in used in order to adaptively choose the non-inferiority threshold $\epsilon$ as follows : if the estimated treatment effect is $\widehat{U}_{Y}$, the significance level $\alpha$ and the desired power to detect a treatment effect based upon the candidate surrogate $S$ is $(1-\beta)$, one may select $\epsilon$ as: 
\begin{equation}
    \epsilon = \max\left\{0, \widehat{u}_{Y} - u^{*}_{\alpha, \beta}\right\}, \label{eps_def}
\end{equation}
where 
\begin{equation*}
    u^{*}_{\alpha, \beta} = \frac{1}{2} - \sqrt{\ddfrac{1-\widehat{\pi}}{4n}} \left[ \Phi^{-1}(\beta) - \Phi^{-1}(1-\alpha) \right].
\end{equation*}

Now, we use similar arguments to derive the variance of $\widehat{\delta} = \widehat{U_{Y}}-\widehat{U_{S}}$. 

\begin{equation*}
\begin{split}
\widehat{\delta} &= \widehat{{U}_{Y}} - \widehat{{U}_{S}} \\
 &= \frac{1}{n}\mathlarger{\sum\limits_{i=1}^{n}G(Y_{i}^{1},Y_{i}^{0})} - \frac{1}{n}\mathlarger{\sum\limits_{i=1}^{n}G(S_{i}^{1},S_{i}^{0})} \\
 &= \frac{1}{n}\mathlarger{\sum\limits_{i=1}^{n} \lbrack G(Y_{i}^{1},Y_{i}^{0})} - G(S_{i}^{1},S_{i}^{0}) \rbrack \\
 &= \frac{1}{n}\mathlarger{\sum\limits_{i=1}^{n}} d_{i}
\end{split}
\end{equation*}

where $d_{i} = G(Y_{i}^{1},Y_{i}^{0}) - G(S_{i}^{1},S_{i}^{0})$. Then, since individuals are independent, 

\begin{equation*}
\begin{split}
Var(\widehat{\delta}) &= Var(\frac{1}{n}\mathlarger{\sum\limits_{i=1}^{n}} d_{i}) \\
                      &= \frac{1}{n^{2}} \mathlarger{\sum\limits_{i=1}^{n}} Var(d_{i}) \\ 
                      & = \frac{n\sigma_{d}^{2}}{n^{2}} \\
                      & = \frac{\sigma_{d}^{2}}{n}
\end{split}
\end{equation*}

such that $\widehat{\delta} \sim \mathcal{N}(\delta, \frac{\sigma_{d}^{2}}{n})$. In practice, we can estimate $\sigma_{d}$ with its sample estimator

$$\widehat{\sigma_{d}}^{2} = \ddfrac{1}{n-1}\mathlarger{\sum\limits_{i = 1}^{n}}(d_{i}-\Bar{d})^{2}$$ where

$$\Bar{d} = \ddfrac{1}{n}\mathlarger{\sum\limits_{i = 1}^{n}}d_{i}$$

such that $\widehat{Var(\delta)} = \ddfrac{\widehat{\sigma_{d}}^{2}}{n}$.

\section{Supporting figures and tables}

\begin{figure}[ht]
\centerline{\includegraphics[height=20pc]{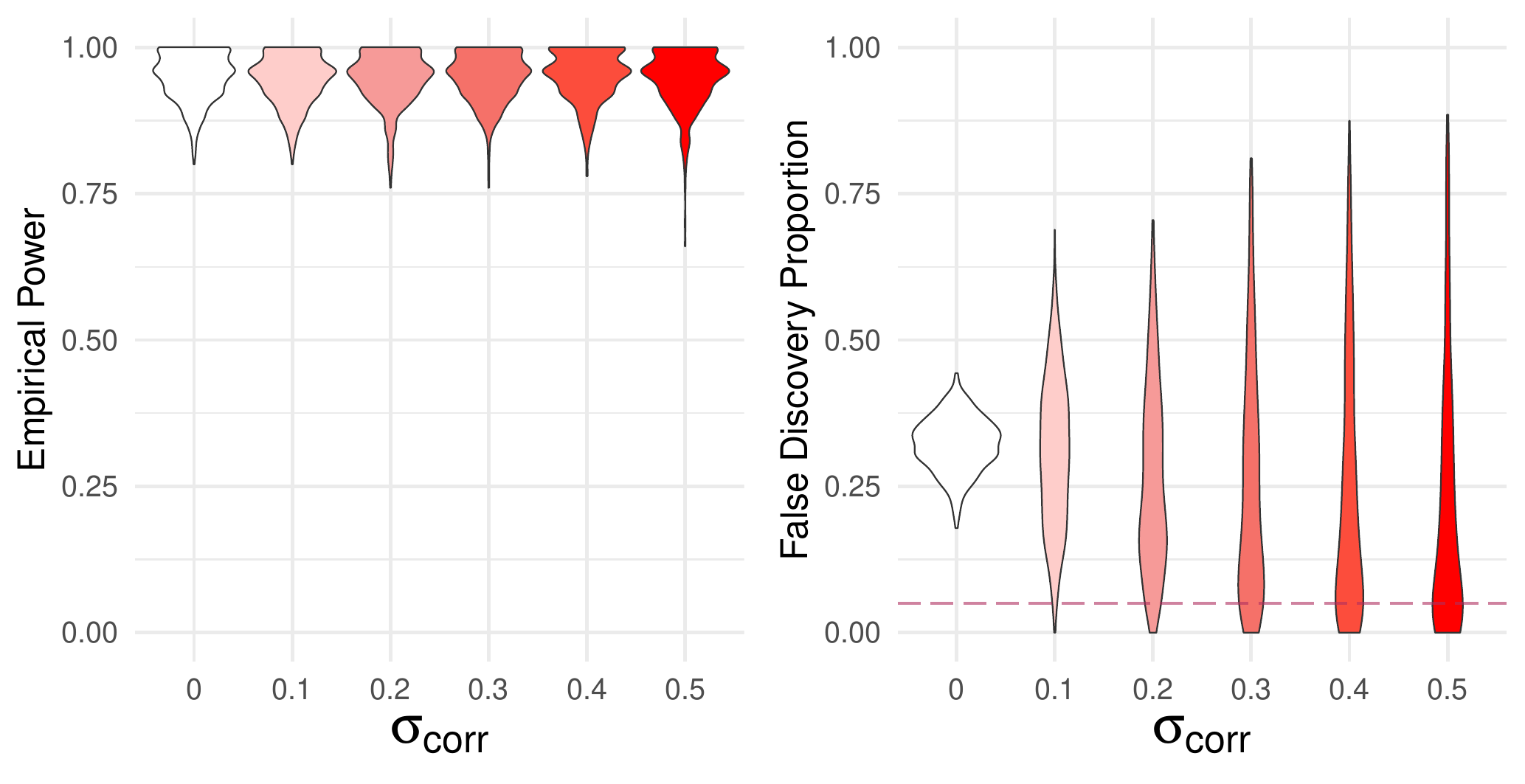}}
\caption{Data generation process 1, scenario 2: violin plots of empirical power (left) and false discovery proportion (right) prior to multiple testing corrections for a fixed sample size $n = 50$ and average surrogate strength $\bar{U_{S}} = 0.9$ for different values of inter-predictor correlation. The nominal significance level $\alpha = 0.05$ is plotted as a dashed purple line on the FDR plot.
\label{figures1}
}
\end{figure}

\begin{figure}[ht]
\centerline{\includegraphics[height=20pc]{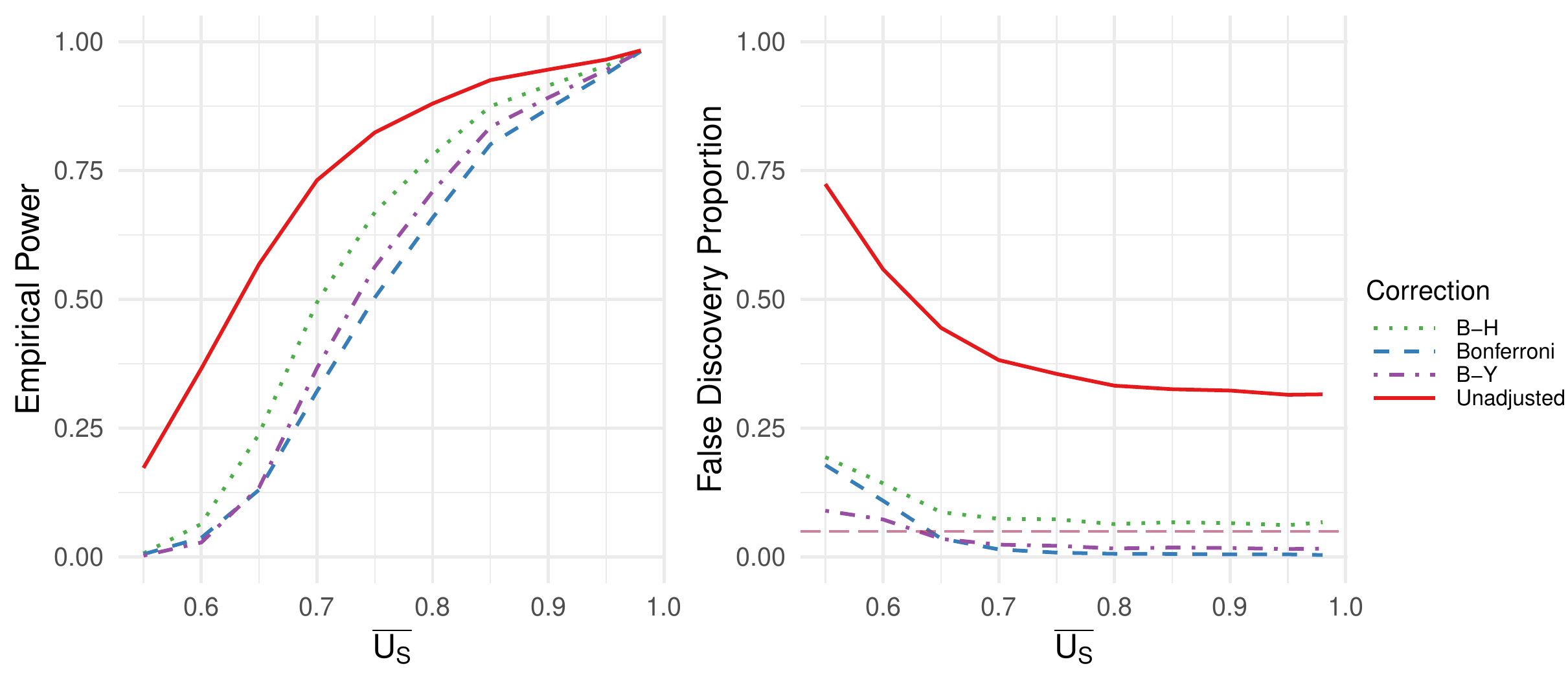}}
\caption{Data generation process 1, scenario 2: Empirical power (left) and false discovery proportion (right) prior to multiple testing corrections as a function of average surrogate strength for different multiple testing corrections (Benjamini-Hochberg, Bonferroni, Benjamini-Yekutieli, Unadjusted) for a fixed sample size $n = 50$. The nominal significance level $\alpha = 0.05$ is plotted as a dashed purple line on the FDR plot.
\label{figures2}
}
\end{figure}

\begin{figure}[ht]
\centerline{\includegraphics[height=20pc]{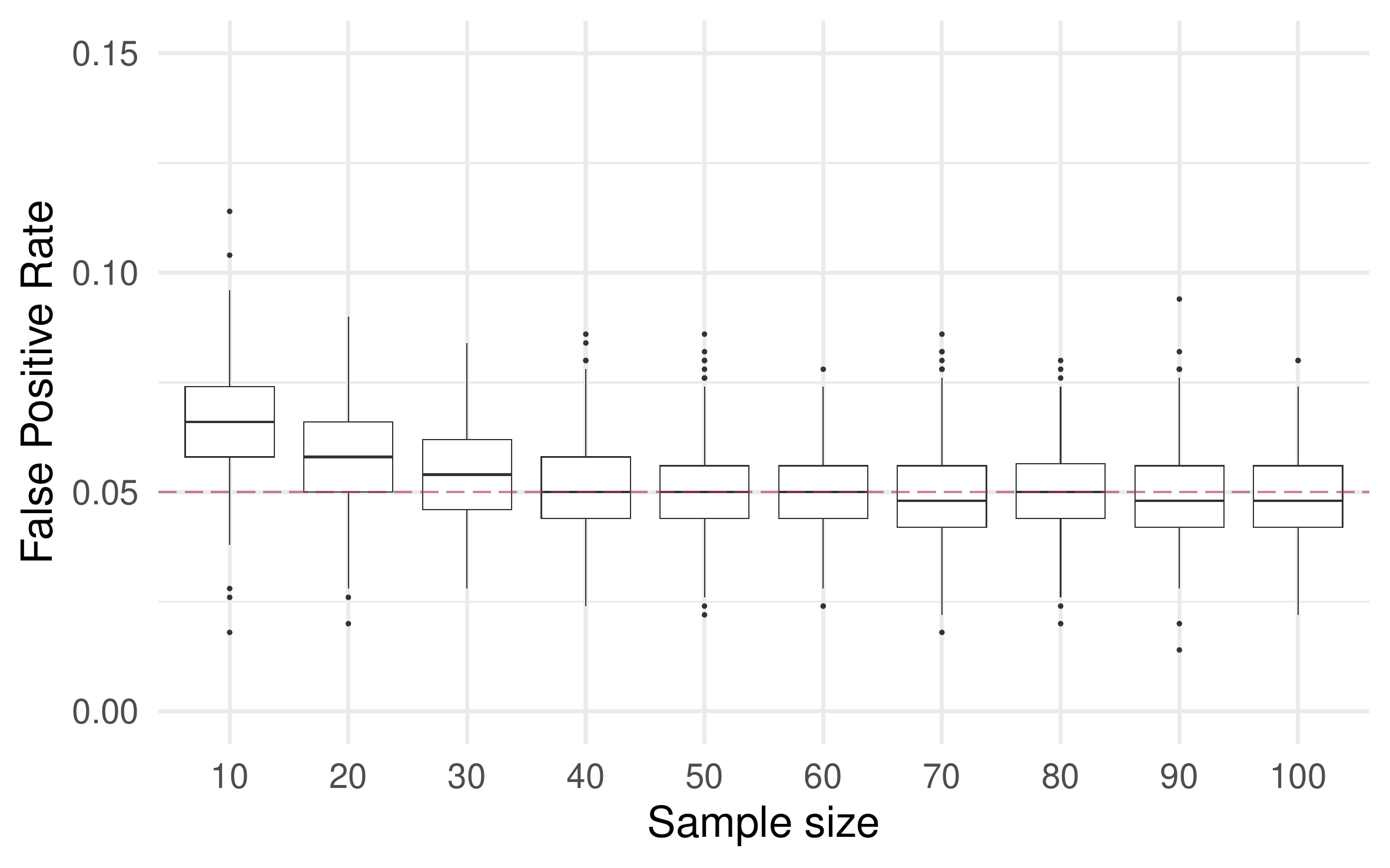}}
\caption{Data generation process 2, scenario 1: boxplots of observed false positive rates against different sample sizes in the uncorrelated setting. The nominal significance level $\alpha = 0.05$ is plotted as a dashed purple line.
\label{figures3}
}
\end{figure}

\begin{figure}[ht]
\centerline{\includegraphics[height=15pc]{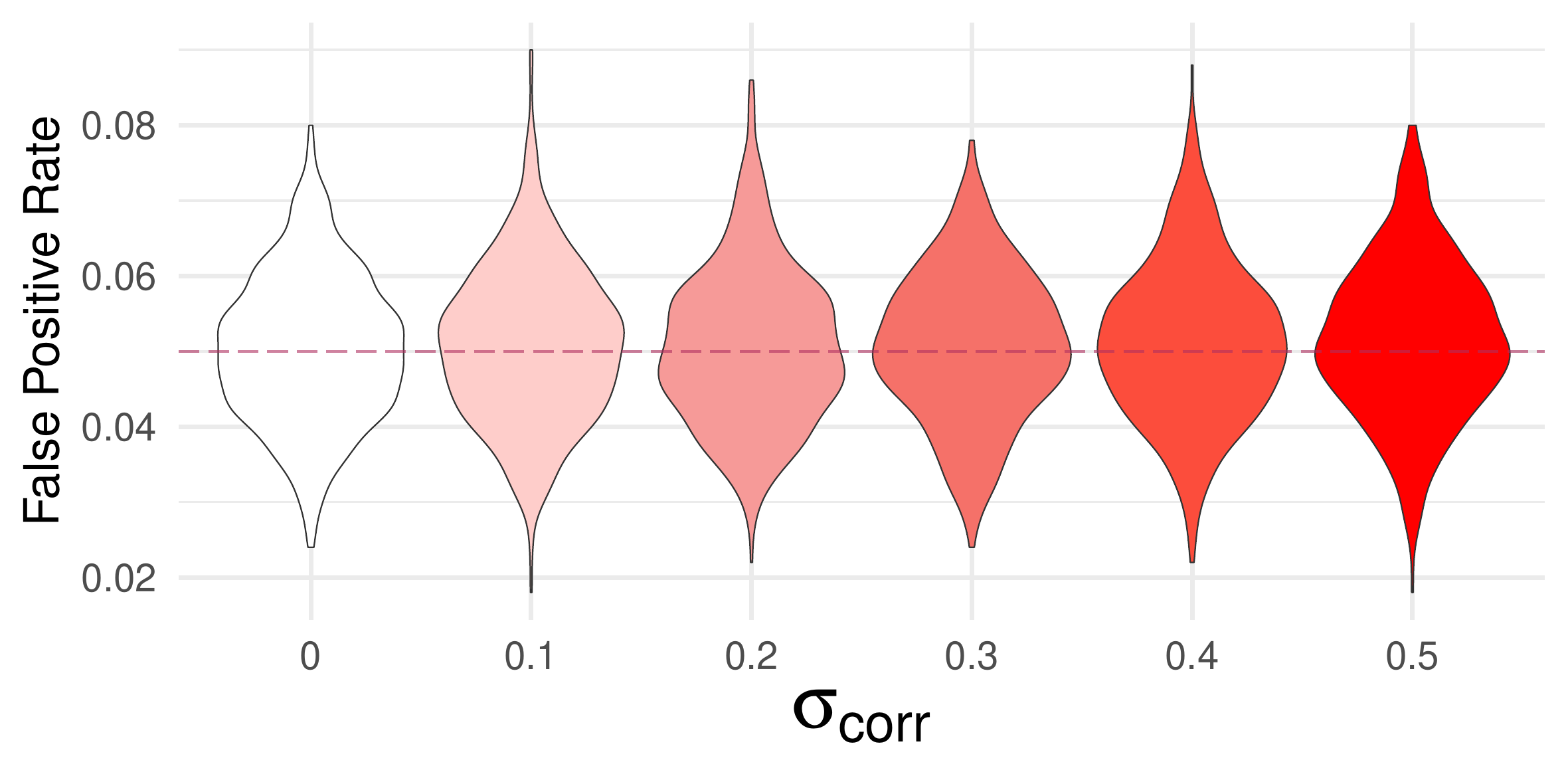}}
\caption{Data generation process 2, scenario 1: violin plots of observed false positive rates against different levels of correlation prior to multiple testing corrections across 500 simulations for a fixed sample size of $n = 50$. The nominal significance level $\alpha = 0.05$ is plotted as a dashed purple line.
\label{figures4}
}
\end{figure}

\begin{figure}[ht]
\centerline{\includegraphics[height=20pc]{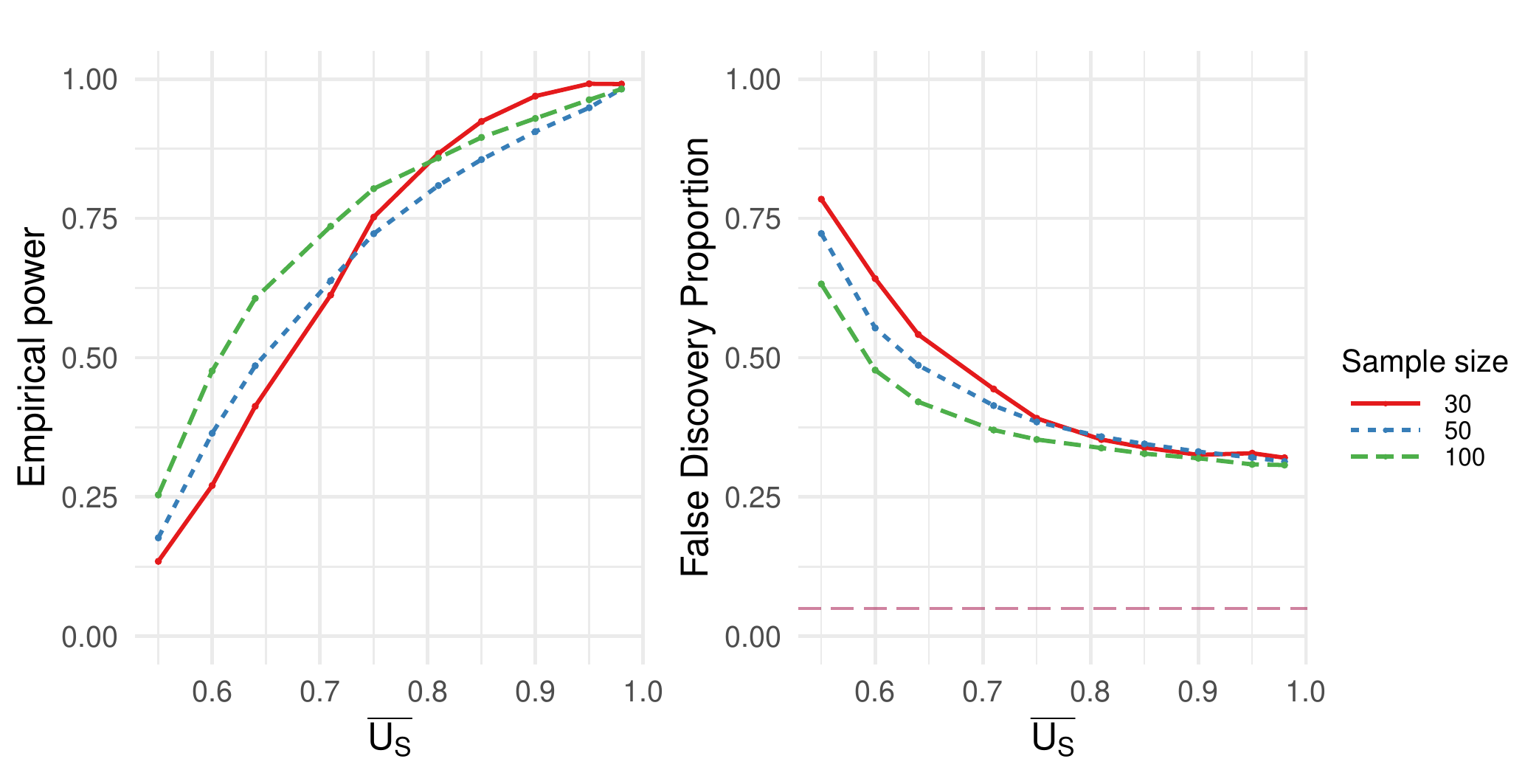}}
\caption{Data generation process 2, scenario 2: empirical power (left) and false discovery proportion (right) prior to multiple testing corrections as a function of average surrogate strength for three different sample sizes. The nominal significance level $\alpha = 0.05$ is plotted as a dashed purple line on the FDR plot.
\label{figures5}
}
\end{figure}

\begin{figure}[ht]
\centerline{\includegraphics[height=20pc]{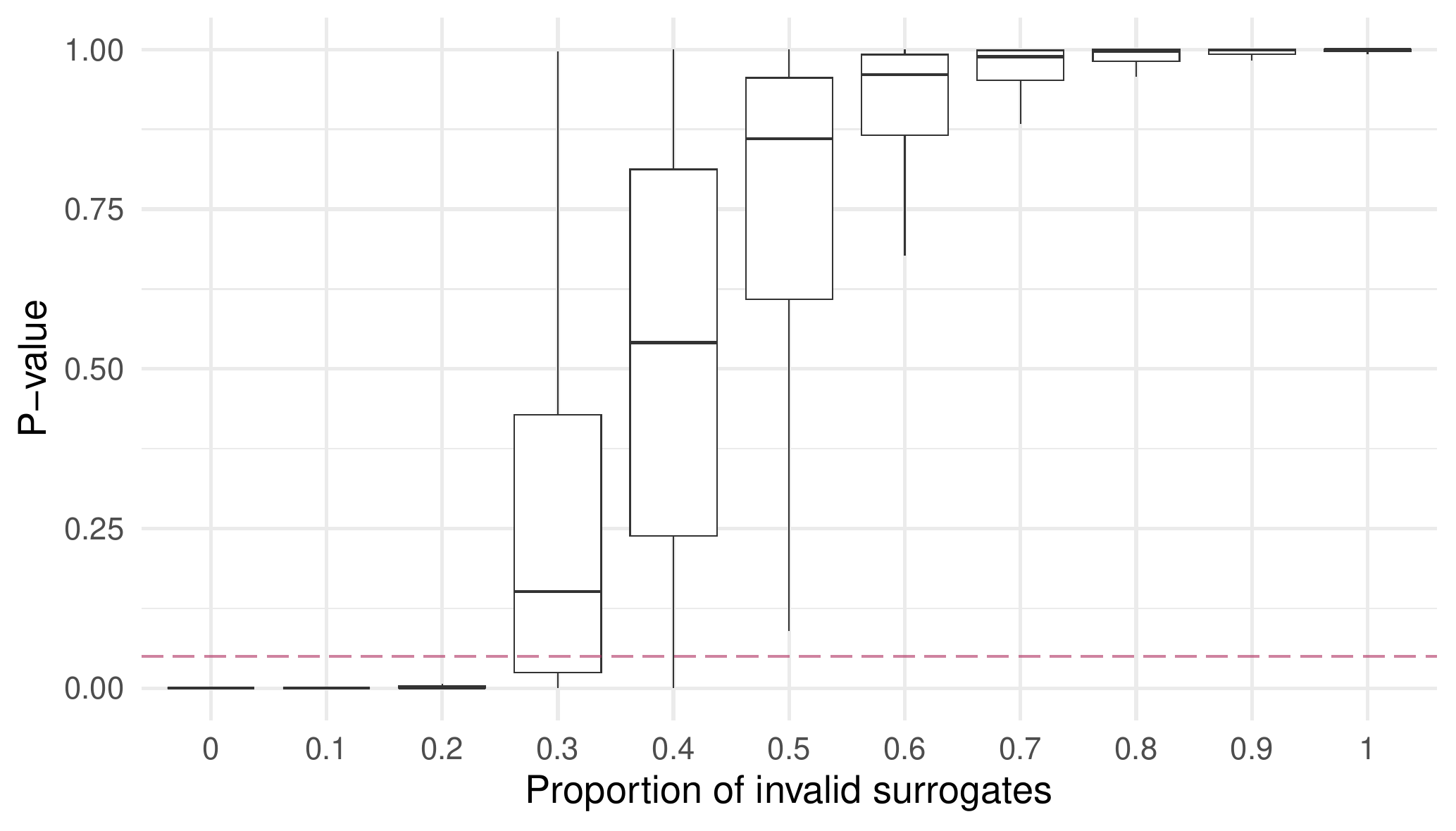}}
\caption{Data generation process 2: The distributions of the p-values in the evaluation step are examined as a function of the false discovery proportion which make up $\widehat{\gamma}_{\mathcal{S}}$, which consists of a combination of $20$ predictors. The sample size is $n = 50$ and the valid surrogate strength is $\widehat{U_{S_{j}}} = 0.9$. The nominal significance level $\alpha = 0.05$ is plotted as a dashed purple line. Desired power for the new surrogate was fixed at $80\%$.
\label{figures6}
}
\end{figure}

\begin{figure}[ht]
\centerline{\includegraphics[height=8.5pc]{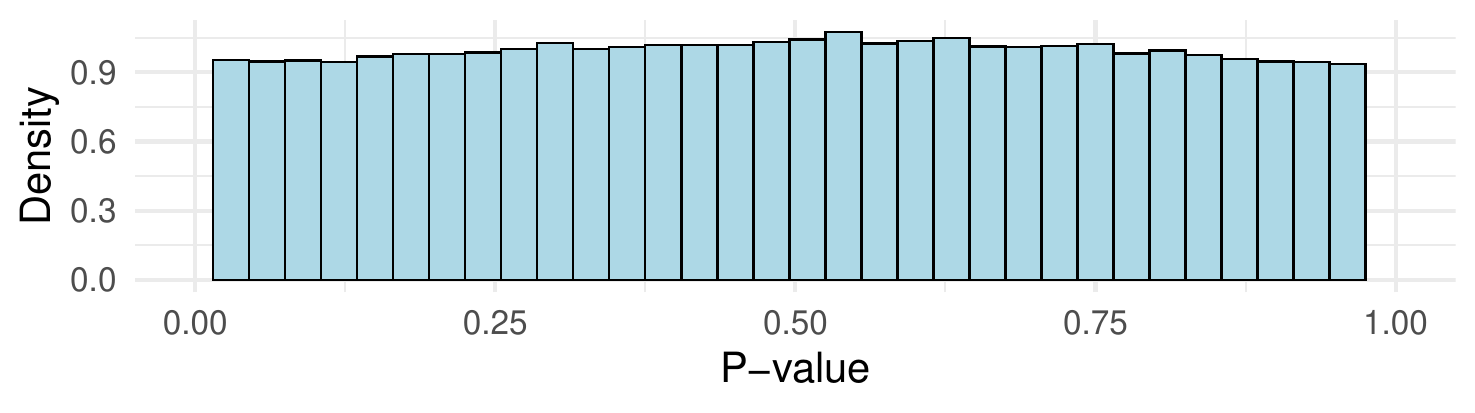}}
\caption{Data generation process 1: distribution of raw p-values under the null hypothesis. The sample size is $n = 50$, the predictors were generated without correlation, and the histogram represents the results across 1000 simulations.\label{figures7}}
\end{figure}

\begin{figure}[ht]
\centerline{\includegraphics[height=20pc]{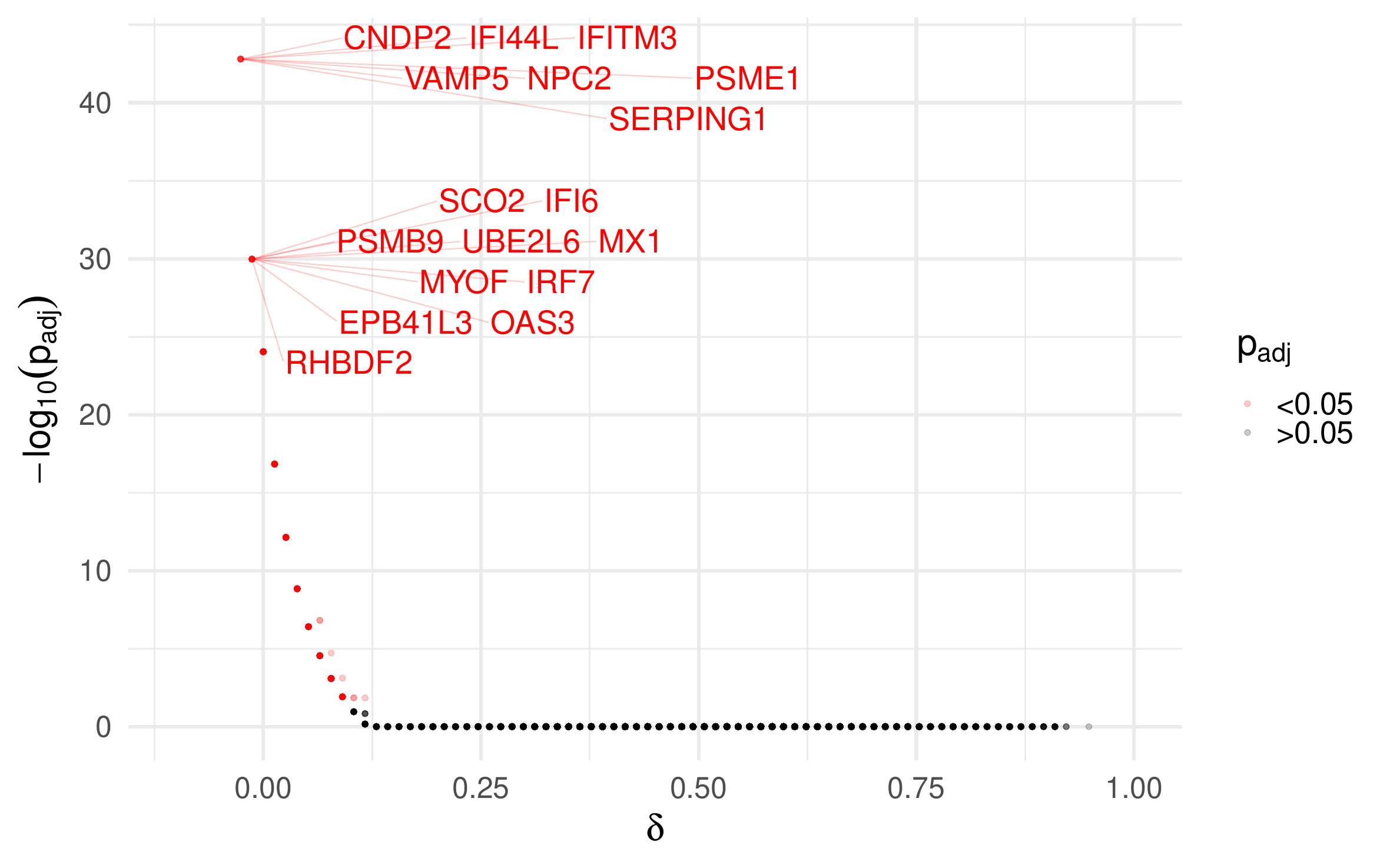}}
\caption{A visual method to select markers to pass the screening stage. The x-axis are the $\delta$ values, and the y-axis is the negative log10 of the adjusted p-value. Markers with a stronger surrogate strength appear towards the top-left of the plot. The 222-genes with an adjusted p-value less than 0.05 are highlighted in red- note that many points are on top of each other due to equivalent p-values resulting from the paired sample test.
\label{figures8}
}
\end{figure}

\begin{figure}[ht]
\centerline{\includegraphics[height=15pc]{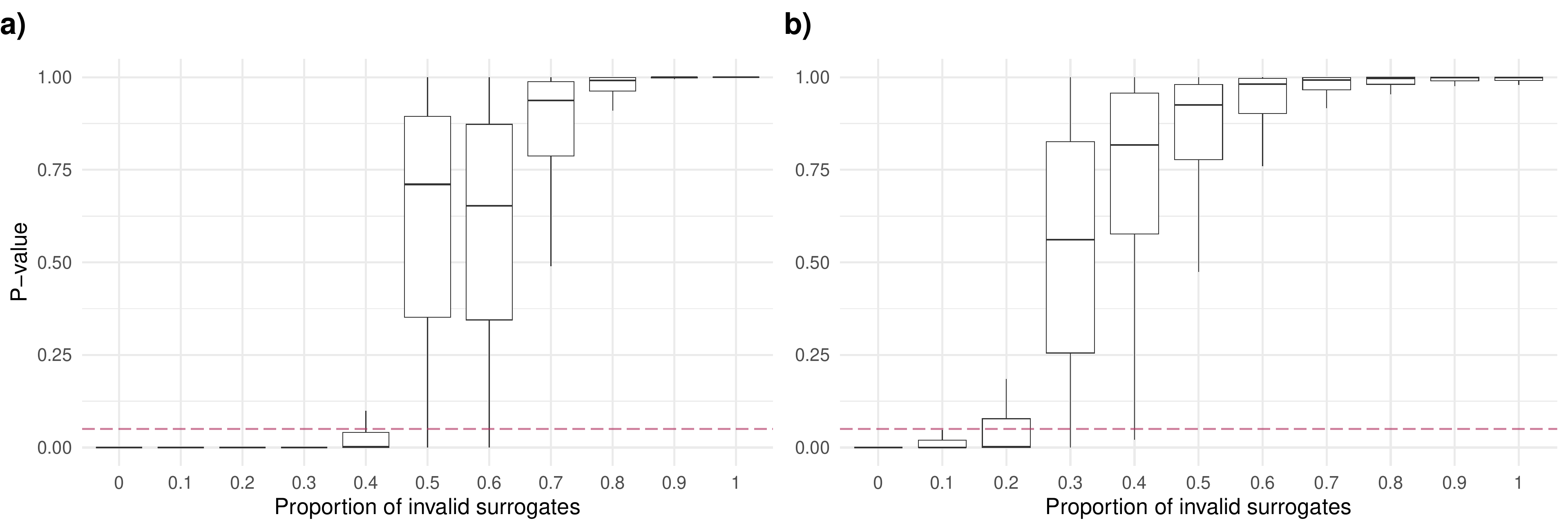}}
\caption{The distributions of the p-values in the evaluation step are examined as a function of the false discovery proportion which make up $\widehat{\gamma}_{\mathcal{S}}$, which consists of a combination of a) 100 predictors and b) 10 predictors. The sample size is $n = 50$ and the valid surrogate strength is $\widehat{U_{S_{j}}} = 0.9$. The nominal significance level $\alpha = 0.05$ is plotted as a dashed purple line. Desired power for the new surrogate was fixed at $80\%$.
\label{figures9}}
\end{figure}

\clearpage

\footnotesize
\begin{longtable}{@{\extracolsep\fill}lllll@{\extracolsep\fill}}
\caption{Screening results from the data application - all genes with adjusted p-values less than 0.05. \label{tables1}}\\
\toprule
\textbf{Gene} & $\boldsymbol{\delta}$ \textbf{(95\% C.I.)} & $\boldsymbol{\sigma_{\delta}}$ & \textbf{Unadjusted p-value} & \textbf{Bonferroni Adjusted p-value} \\
\midrule
\endfirsthead

\toprule
\textbf{Gene} & $\boldsymbol{\delta}$ \textbf{(95\% C.I.)} & $\boldsymbol{\sigma_{\delta}}$ & \textbf{Unadjusted p-value} & \textbf{Bonferroni Adjusted p-value} \\
\midrule
\endhead
\bottomrule
\endfoot
  CNDP2 & -0.026 (-0.056, 0.004) & 0.018 & 1.6e-47 & 1.6e-43 \\ 
  IFI44L & -0.026 (-0.056, 0.004) & 0.018 & 1.6e-47 & 1.6e-43 \\ 
  IFITM3 & -0.026 (-0.056, 0.004) & 0.018 & 1.6e-47 & 1.6e-43 \\ 
  NPC2 & -0.026 (-0.056, 0.004) & 0.018 & 1.6e-47 & 1.6e-43 \\ 
  PSME1 & -0.026 (-0.056, 0.004) & 0.018 & 1.6e-47 & 1.6e-43 \\ 
  SERPING1 & -0.026 (-0.056, 0.004) & 0.018 & 1.6e-47 & 1.6e-43 \\ 
  VAMP5 & -0.026 (-0.056, 0.004) & 0.018 & 1.6e-47 & 1.6e-43 \\ 
  EPB41L3 & -0.013 (-0.05, 0.024) & 0.023 & 1.1e-34 & 1.1e-30 \\ 
  IFI6 & -0.013 (-0.05, 0.024) & 0.023 & 1.1e-34 & 1.1e-30 \\ 
  IRF7 & -0.013 (-0.05, 0.024) & 0.023 & 1.1e-34 & 1.1e-30 \\ 
  MX1 & -0.013 (-0.05, 0.024) & 0.023 & 1.1e-34 & 1.1e-30 \\ 
  MYOF & -0.013 (-0.05, 0.024) & 0.023 & 1.1e-34 & 1.1e-30 \\ 
  OAS3 & -0.013 (-0.05, 0.024) & 0.023 & 1.1e-34 & 1.1e-30 \\ 
  PSMB9 & -0.013 (-0.05, 0.024) & 0.023 & 1.1e-34 & 1.1e-30 \\ 
  RHBDF2 & -0.013 (-0.05, 0.024) & 0.023 & 1.1e-34 & 1.1e-30 \\ 
  SCO2 & -0.013 (-0.05, 0.024) & 0.023 & 1.1e-34 & 1.1e-30 \\ 
  UBE2L6 & -0.013 (-0.05, 0.024) & 0.023 & 1.1e-34 & 1.1e-30 \\ 
  WARS1 & -0.013 (-0.05, 0.024) & 0.023 & 1.1e-34 & 1.1e-30 \\ 
  ADAP2 & 0 (-0.043, 0.043) & 0.026 & 9.1e-29 & 9.2e-25 \\ 
  BST2 & 0 (-0.043, 0.043) & 0.026 & 9.1e-29 & 9.2e-25 \\ 
  CEACAM1 & 0 (-0.043, 0.043) & 0.026 & 9.1e-29 & 9.2e-25 \\ 
  CYBB & 0 (-0.043, 0.043) & 0.026 & 9.1e-29 & 9.2e-25 \\ 
  HERC5 & 0 (-0.043, 0.043) & 0.026 & 9.1e-29 & 9.2e-25 \\ 
  IFI35 & 0 (-0.043, 0.043) & 0.026 & 9.1e-29 & 9.2e-25 \\ 
  IFIH1 & 0 (-0.043, 0.043) & 0.026 & 9.1e-29 & 9.2e-25 \\ 
  IFITM1 & 0 (-0.043, 0.043) & 0.026 & 9.1e-29 & 9.2e-25 \\ 
  LY6E & 0 (-0.043, 0.043) & 0.026 & 9.1e-29 & 9.2e-25 \\ 
  MICB & 0 (-0.043, 0.043) & 0.026 & 9.1e-29 & 9.2e-25 \\ 
  NAGK & 0 (-0.043, 0.043) & 0.026 & 9.1e-29 & 9.2e-25 \\ 
  OAS1 & 0 (-0.043, 0.043) & 0.026 & 9.1e-29 & 9.2e-25 \\ 
  OASL & 0 (-0.043, 0.043) & 0.026 & 9.1e-29 & 9.2e-25 \\ 
  P2RX7 & 0 (-0.043, 0.043) & 0.026 & 9.1e-29 & 9.2e-25 \\ 
  PSMB10 & 0 (-0.043, 0.043) & 0.026 & 9.1e-29 & 9.2e-25 \\ 
  RSAD2 & 0 (-0.043, 0.043) & 0.026 & 9.1e-29 & 9.2e-25 \\ 
  RTP4 & 0 (-0.043, 0.043) & 0.026 & 9.1e-29 & 9.2e-25 \\ 
  SCARB2 & 0 (-0.043, 0.043) & 0.026 & 9.1e-29 & 9.2e-25 \\ 
  SQOR & 0 (-0.043, 0.043) & 0.026 & 9.1e-29 & 9.2e-25 \\ 
  STAT2 & 0 (-0.043, 0.043) & 0.026 & 9.1e-29 & 9.2e-25 \\ 
  TLR7 & 0 (-0.043, 0.043) & 0.026 & 9.1e-29 & 9.2e-25 \\ 
  TRIM21 & 0 (-0.043, 0.043) & 0.026 & 9.1e-29 & 9.2e-25 \\ 
  TYMP & 0 (-0.043, 0.043) & 0.026 & 9.1e-29 & 9.2e-25 \\ 
  XAF1 & 0 (-0.043, 0.043) & 0.026 & 9.1e-29 & 9.2e-25 \\ 
  ADAR & 0.013 (-0.035, 0.061) & 0.029 & 1.5e-21 & 1.5e-17 \\ 
  AFF1 & 0.013 (-0.035, 0.061) & 0.029 & 1.5e-21 & 1.5e-17 \\ 
  ATP1B3 & 0.013 (-0.035, 0.061) & 0.029 & 1.5e-21 & 1.5e-17 \\ 
  DHRS9 & 0.013 (-0.035, 0.061) & 0.029 & 1.5e-21 & 1.5e-17 \\ 
  EIF2AK2 & 0.013 (-0.035, 0.061) & 0.029 & 1.5e-21 & 1.5e-17 \\ 
  GBP1 & 0.013 (-0.035, 0.061) & 0.029 & 1.5e-21 & 1.5e-17 \\ 
  GBP2 & 0.013 (-0.035, 0.061) & 0.029 & 1.5e-21 & 1.5e-17 \\ 
  GCH1 & 0.013 (-0.035, 0.061) & 0.029 & 1.5e-21 & 1.5e-17 \\ 
  GNS & 0.013 (-0.035, 0.061) & 0.029 & 1.5e-21 & 1.5e-17 \\ 
  GSDMD & 0.013 (-0.035, 0.061) & 0.029 & 1.5e-21 & 1.5e-17 \\ 
  IFIT3 & 0.013 (-0.035, 0.061) & 0.029 & 1.5e-21 & 1.5e-17 \\ 
  MAFB & 0.013 (-0.035, 0.061) & 0.029 & 1.5e-21 & 1.5e-17 \\ 
  MT2A & 0.013 (-0.035, 0.061) & 0.029 & 1.5e-21 & 1.5e-17 \\ 
  NOD2 & 0.013 (-0.035, 0.061) & 0.029 & 1.5e-21 & 1.5e-17 \\ 
  OAS2 & 0.013 (-0.035, 0.061) & 0.029 & 1.5e-21 & 1.5e-17 \\ 
  RBCK1 & 0.013 (-0.035, 0.061) & 0.029 & 1.5e-21 & 1.5e-17 \\ 
  SHTN1 & 0.013 (-0.035, 0.061) & 0.029 & 1.5e-21 & 1.5e-17 \\ 
  SRBD1 & 0.013 (-0.035, 0.061) & 0.029 & 1.5e-21 & 1.5e-17 \\ 
  STAT1 & 0.013 (-0.035, 0.061) & 0.029 & 1.5e-21 & 1.5e-17 \\ 
  TBC1D2B & 0.013 (-0.035, 0.061) & 0.029 & 1.5e-21 & 1.5e-17 \\ 
  TNF & 0.013 (-0.035, 0.061) & 0.029 & 1.5e-21 & 1.5e-17 \\ 
  TNFAIP6 & 0.013 (-0.035, 0.061) & 0.029 & 1.5e-21 & 1.5e-17 \\ 
  AKR1A1 & 0.026 (-0.026, 0.078) & 0.032 & 7.3e-17 & 7.3e-13 \\ 
  ALDH1A1 & 0.026 (-0.026, 0.078) & 0.032 & 7.3e-17 & 7.3e-13 \\ 
  ALDH2 & 0.026 (-0.026, 0.078) & 0.032 & 7.3e-17 & 7.3e-13 \\ 
  ARSB & 0.026 (-0.026, 0.078) & 0.032 & 7.3e-17 & 7.3e-13 \\ 
  ATF5 & 0.026 (-0.026, 0.078) & 0.032 & 7.3e-17 & 7.3e-13 \\ 
  CALCOCO2 & 0.026 (-0.026, 0.078) & 0.032 & 7.3e-17 & 7.3e-13 \\ 
  DDX58 & 0.026 (-0.026, 0.078) & 0.032 & 7.3e-17 & 7.3e-13 \\ 
  DENND1A & 0.026 (-0.026, 0.078) & 0.032 & 7.3e-17 & 7.3e-13 \\ 
  DRAP1 & 0.026 (-0.026, 0.078) & 0.032 & 7.3e-17 & 7.3e-13 \\ 
  HLA\_F & 0.026 (-0.026, 0.078) & 0.032 & 7.3e-17 & 7.3e-13 \\ 
  IFI44 & 0.026 (-0.026, 0.078) & 0.032 & 7.3e-17 & 7.3e-13 \\ 
  IFIT2 & 0.026 (-0.026, 0.078) & 0.032 & 7.3e-17 & 7.3e-13 \\ 
  IRF9 & 0.026 (-0.026, 0.078) & 0.032 & 7.3e-17 & 7.3e-13 \\ 
  KYNU & 0.026 (-0.026, 0.078) & 0.032 & 7.3e-17 & 7.3e-13 \\ 
  LHFPL2 & 0.026 (-0.026, 0.078) & 0.032 & 7.3e-17 & 7.3e-13 \\ 
  MSRB2 & 0.026 (-0.026, 0.078) & 0.032 & 7.3e-17 & 7.3e-13 \\ 
  MTMR11 & 0.026 (-0.026, 0.078) & 0.032 & 7.3e-17 & 7.3e-13 \\ 
  PLSCR1 & 0.026 (-0.026, 0.078) & 0.032 & 7.3e-17 & 7.3e-13 \\ 
  SLC2A6 & 0.026 (-0.026, 0.078) & 0.032 & 7.3e-17 & 7.3e-13 \\ 
  SORT1 & 0.026 (-0.026, 0.078) & 0.032 & 7.3e-17 & 7.3e-13 \\ 
  SP110 & 0.026 (-0.026, 0.078) & 0.032 & 7.3e-17 & 7.3e-13 \\ 
  SP140 & 0.026 (-0.026, 0.078) & 0.032 & 7.3e-17 & 7.3e-13 \\ 
  STX11 & 0.026 (-0.026, 0.078) & 0.032 & 7.3e-17 & 7.3e-13 \\ 
  TDRD7 & 0.026 (-0.026, 0.078) & 0.032 & 7.3e-17 & 7.3e-13 \\ 
  TENT5A & 0.026 (-0.026, 0.078) & 0.032 & 7.3e-17 & 7.3e-13 \\ 
  TRAFD1 & 0.026 (-0.026, 0.078) & 0.032 & 7.3e-17 & 7.3e-13 \\ 
  UNC93B1 & 0.026 (-0.026, 0.078) & 0.032 & 7.3e-17 & 7.3e-13 \\ 
  ASGR2 & 0.039 (-0.017, 0.095) & 0.034 & 1.4e-13 & 1.5e-09 \\ 
  ATOX1 & 0.039 (-0.017, 0.095) & 0.034 & 1.4e-13 & 1.5e-09 \\ 
  C1QB & 0.039 (-0.017, 0.095) & 0.034 & 1.4e-13 & 1.5e-09 \\ 
  CD300A & 0.039 (-0.017, 0.095) & 0.034 & 1.4e-13 & 1.5e-09 \\ 
  DRAM1 & 0.039 (-0.017, 0.095) & 0.034 & 1.4e-13 & 1.5e-09 \\ 
  DUSP3 & 0.039 (-0.017, 0.095) & 0.034 & 1.4e-13 & 1.5e-09 \\ 
  DUSP5 & 0.039 (-0.017, 0.095) & 0.034 & 1.4e-13 & 1.5e-09 \\ 
  EMILIN2 & 0.039 (-0.017, 0.095) & 0.034 & 1.4e-13 & 1.5e-09 \\ 
  IRF1 & 0.039 (-0.017, 0.095) & 0.034 & 1.4e-13 & 1.5e-09 \\ 
  ISG20 & 0.039 (-0.017, 0.095) & 0.034 & 1.4e-13 & 1.5e-09 \\ 
  MX2 & 0.039 (-0.017, 0.095) & 0.034 & 1.4e-13 & 1.5e-09 \\ 
  P2RY14 & 0.039 (-0.017, 0.095) & 0.034 & 1.4e-13 & 1.5e-09 \\ 
  PANK2 & 0.039 (-0.017, 0.095) & 0.034 & 1.4e-13 & 1.5e-09 \\ 
  PARP12 & 0.039 (-0.017, 0.095) & 0.034 & 1.4e-13 & 1.5e-09 \\ 
  PLEK & 0.039 (-0.017, 0.095) & 0.034 & 1.4e-13 & 1.5e-09 \\ 
  PLEKHO1 & 0.039 (-0.017, 0.095) & 0.034 & 1.4e-13 & 1.5e-09 \\ 
  PSMB2 & 0.039 (-0.017, 0.095) & 0.034 & 1.4e-13 & 1.5e-09 \\ 
  PSTPIP2 & 0.039 (-0.017, 0.095) & 0.034 & 1.4e-13 & 1.5e-09 \\ 
  SAMD4A & 0.039 (-0.017, 0.095) & 0.034 & 1.4e-13 & 1.5e-09 \\ 
  SLC6A12 & 0.039 (-0.017, 0.095) & 0.034 & 1.4e-13 & 1.5e-09 \\ 
  SNTB1 & 0.039 (-0.017, 0.095) & 0.034 & 1.4e-13 & 1.5e-09 \\ 
  SPATS2L & 0.039 (-0.017, 0.095) & 0.034 & 1.4e-13 & 1.5e-09 \\ 
  TFIP11 & 0.039 (-0.017, 0.095) & 0.034 & 1.4e-13 & 1.5e-09 \\ 
  TIMM10 & 0.039 (-0.017, 0.095) & 0.034 & 1.4e-13 & 1.5e-09 \\ 
  TNFAIP2 & 0.039 (-0.017, 0.095) & 0.034 & 1.4e-13 & 1.5e-09 \\ 
  TRIM5 & 0.039 (-0.017, 0.095) & 0.034 & 1.4e-13 & 1.5e-09 \\ 
  FGR & 0.065 (0.009, 0.12) & 0.034 & 1.5e-11 & 1.5e-07 \\ 
  HERC6 & 0.065 (0.009, 0.12) & 0.034 & 1.5e-11 & 1.5e-07 \\ 
  AIM2 & 0.052 (-0.008, 0.112) & 0.036 & 3.9e-11 & 3.9e-07 \\ 
  ANKFY1 & 0.052 (-0.008, 0.112) & 0.036 & 3.9e-11 & 3.9e-07 \\ 
  ATF3 & 0.052 (-0.008, 0.112) & 0.036 & 3.9e-11 & 3.9e-07 \\ 
  BLVRA & 0.052 (-0.008, 0.112) & 0.036 & 3.9e-11 & 3.9e-07 \\ 
  CTNNA1 & 0.052 (-0.008, 0.112) & 0.036 & 3.9e-11 & 3.9e-07 \\ 
  CXCL10 & 0.052 (-0.008, 0.112) & 0.036 & 3.9e-11 & 3.9e-07 \\ 
  DDX60 & 0.052 (-0.008, 0.112) & 0.036 & 3.9e-11 & 3.9e-07 \\ 
  DHX58 & 0.052 (-0.008, 0.112) & 0.036 & 3.9e-11 & 3.9e-07 \\ 
  DPYD & 0.052 (-0.008, 0.112) & 0.036 & 3.9e-11 & 3.9e-07 \\ 
  FAM111A & 0.052 (-0.008, 0.112) & 0.036 & 3.9e-11 & 3.9e-07 \\ 
  HLA\_DMA & 0.052 (-0.008, 0.112) & 0.036 & 3.9e-11 & 3.9e-07 \\ 
  IFIT1 & 0.052 (-0.008, 0.112) & 0.036 & 3.9e-11 & 3.9e-07 \\ 
  IRF2 & 0.052 (-0.008, 0.112) & 0.036 & 3.9e-11 & 3.9e-07 \\ 
  KCNJ2 & 0.052 (-0.008, 0.112) & 0.036 & 3.9e-11 & 3.9e-07 \\ 
  LILRB2 & 0.052 (-0.008, 0.112) & 0.036 & 3.9e-11 & 3.9e-07 \\ 
  NFKBIE & 0.052 (-0.008, 0.112) & 0.036 & 3.9e-11 & 3.9e-07 \\ 
  PHF11 & 0.052 (-0.008, 0.112) & 0.036 & 3.9e-11 & 3.9e-07 \\ 
  PLAGL1 & 0.052 (-0.008, 0.112) & 0.036 & 3.9e-11 & 3.9e-07 \\ 
  PSMB8 & 0.052 (-0.008, 0.112) & 0.036 & 3.9e-11 & 3.9e-07 \\ 
  SRC & 0.052 (-0.008, 0.112) & 0.036 & 3.9e-11 & 3.9e-07 \\ 
  TAPBPL & 0.052 (-0.008, 0.112) & 0.036 & 3.9e-11 & 3.9e-07 \\ 
  TRIM22 & 0.052 (-0.008, 0.112) & 0.036 & 3.9e-11 & 3.9e-07 \\ 
  DECR1 & 0.078 (0.019, 0.137) & 0.036 & 1.9e-09 & 1.9e-05 \\ 
  ACTA2 & 0.065 (0.002, 0.128) & 0.039 & 2.8e-09 & 2.8e-05 \\ 
  CD300C & 0.065 (0.002, 0.128) & 0.039 & 2.8e-09 & 2.8e-05 \\ 
  CTRL & 0.065 (0.002, 0.128) & 0.039 & 2.8e-09 & 2.8e-05 \\ 
  CTSS & 0.065 (0.002, 0.128) & 0.039 & 2.8e-09 & 2.8e-05 \\ 
  FCN1 & 0.065 (0.002, 0.128) & 0.039 & 2.8e-09 & 2.8e-05 \\ 
  HEBP1 & 0.065 (0.002, 0.128) & 0.039 & 2.8e-09 & 2.8e-05 \\ 
  HLA\_DMB & 0.065 (0.002, 0.128) & 0.039 & 2.8e-09 & 2.8e-05 \\ 
  HLA\_DPA1 & 0.065 (0.002, 0.128) & 0.039 & 2.8e-09 & 2.8e-05 \\ 
  HLA\_DRA & 0.065 (0.002, 0.128) & 0.039 & 2.8e-09 & 2.8e-05 \\ 
  IFI16 & 0.065 (0.002, 0.128) & 0.039 & 2.8e-09 & 2.8e-05 \\ 
  IL15 & 0.065 (0.002, 0.128) & 0.039 & 2.8e-09 & 2.8e-05 \\ 
  KCNMB1 & 0.065 (0.002, 0.128) & 0.039 & 2.8e-09 & 2.8e-05 \\ 
  KLF4 & 0.065 (0.002, 0.128) & 0.039 & 2.8e-09 & 2.8e-05 \\ 
  MICU1 & 0.065 (0.002, 0.128) & 0.039 & 2.8e-09 & 2.8e-05 \\ 
  NUCB1 & 0.065 (0.002, 0.128) & 0.039 & 2.8e-09 & 2.8e-05 \\ 
  OGFR & 0.065 (0.002, 0.128) & 0.039 & 2.8e-09 & 2.8e-05 \\ 
  PLAAT4 & 0.065 (0.002, 0.128) & 0.039 & 2.8e-09 & 2.8e-05 \\ 
  PLAGL2 & 0.065 (0.002, 0.128) & 0.039 & 2.8e-09 & 2.8e-05 \\ 
  PSMA5 & 0.065 (0.002, 0.128) & 0.039 & 2.8e-09 & 2.8e-05 \\ 
  REC8 & 0.065 (0.002, 0.128) & 0.039 & 2.8e-09 & 2.8e-05 \\ 
  TCN2 & 0.065 (0.002, 0.128) & 0.039 & 2.8e-09 & 2.8e-05 \\ 
  TMEM140 & 0.065 (0.002, 0.128) & 0.039 & 2.8e-09 & 2.8e-05 \\ 
  TNS3 & 0.065 (0.002, 0.128) & 0.039 & 2.8e-09 & 2.8e-05 \\ 
  DPYSL2 & 0.091 (0.029, 0.153) & 0.038 & 7.7e-08 & 7.7e-04 \\ 
  APOBEC3G & 0.078 (0.012, 0.144) & 0.040 & 8.1e-08 & 8.2e-04 \\ 
  ASCL2 & 0.078 (0.012, 0.144) & 0.040 & 8.1e-08 & 8.2e-04 \\ 
  ASGR1 & 0.078 (0.012, 0.144) & 0.040 & 8.1e-08 & 8.2e-04 \\ 
  BTN3A3 & 0.078 (0.012, 0.144) & 0.040 & 8.1e-08 & 8.2e-04 \\ 
  CD74 & 0.078 (0.012, 0.144) & 0.040 & 8.1e-08 & 8.2e-04 \\ 
  CDC42EP2 & 0.078 (0.012, 0.144) & 0.040 & 8.1e-08 & 8.2e-04 \\ 
  CTSL & 0.078 (0.012, 0.144) & 0.040 & 8.1e-08 & 8.2e-04 \\ 
  CUL1 & 0.078 (0.012, 0.144) & 0.040 & 8.1e-08 & 8.2e-04 \\ 
  DMXL2 & 0.078 (0.012, 0.144) & 0.040 & 8.1e-08 & 8.2e-04 \\ 
  ETV6 & 0.078 (0.012, 0.144) & 0.040 & 8.1e-08 & 8.2e-04 \\ 
  FAR2 & 0.078 (0.012, 0.144) & 0.040 & 8.1e-08 & 8.2e-04 \\ 
  FFAR2 & 0.078 (0.012, 0.144) & 0.040 & 8.1e-08 & 8.2e-04 \\ 
  FYB1 & 0.078 (0.012, 0.144) & 0.040 & 8.1e-08 & 8.2e-04 \\ 
  GADD45B & 0.078 (0.012, 0.144) & 0.040 & 8.1e-08 & 8.2e-04 \\ 
  GAS6 & 0.078 (0.012, 0.144) & 0.040 & 8.1e-08 & 8.2e-04 \\ 
  GSTK1 & 0.078 (0.012, 0.144) & 0.040 & 8.1e-08 & 8.2e-04 \\ 
  IL12RB1 & 0.078 (0.012, 0.144) & 0.040 & 8.1e-08 & 8.2e-04 \\ 
  ILK & 0.078 (0.012, 0.144) & 0.040 & 8.1e-08 & 8.2e-04 \\ 
  IRF5 & 0.078 (0.012, 0.144) & 0.040 & 8.1e-08 & 8.2e-04 \\ 
  KPNB1 & 0.078 (0.012, 0.144) & 0.040 & 8.1e-08 & 8.2e-04 \\ 
  LILRB4 & 0.078 (0.012, 0.144) & 0.040 & 8.1e-08 & 8.2e-04 \\ 
  MRPL44 & 0.078 (0.012, 0.144) & 0.040 & 8.1e-08 & 8.2e-04 \\ 
  MYD88 & 0.078 (0.012, 0.144) & 0.040 & 8.1e-08 & 8.2e-04 \\ 
  PARP3 & 0.078 (0.012, 0.144) & 0.040 & 8.1e-08 & 8.2e-04 \\ 
  PML & 0.078 (0.012, 0.144) & 0.040 & 8.1e-08 & 8.2e-04 \\ 
  RIPK2 & 0.078 (0.012, 0.144) & 0.040 & 8.1e-08 & 8.2e-04 \\ 
  RNF114 & 0.078 (0.012, 0.144) & 0.040 & 8.1e-08 & 8.2e-04 \\ 
  RRAS & 0.078 (0.012, 0.144) & 0.040 & 8.1e-08 & 8.2e-04 \\ 
  SCPEP1 & 0.078 (0.012, 0.144) & 0.040 & 8.1e-08 & 8.2e-04 \\ 
  SEC24D & 0.078 (0.012, 0.144) & 0.040 & 8.1e-08 & 8.2e-04 \\ 
  SLC20A1 & 0.078 (0.012, 0.144) & 0.040 & 8.1e-08 & 8.2e-04 \\ 
  SLC27A3 & 0.078 (0.012, 0.144) & 0.040 & 8.1e-08 & 8.2e-04 \\ 
  SLC7A7 & 0.078 (0.012, 0.144) & 0.040 & 8.1e-08 & 8.2e-04 \\ 
  TAPBP & 0.078 (0.012, 0.144) & 0.040 & 8.1e-08 & 8.2e-04 \\ 
  TOR1B & 0.078 (0.012, 0.144) & 0.040 & 8.1e-08 & 8.2e-04 \\ 
  ACOT9 & 0.091 (0.022, 0.16) & 0.042 & 1.2e-06 & 1.2e-02 \\ 
  ACSL5 & 0.091 (0.022, 0.16) & 0.042 & 1.2e-06 & 1.2e-02 \\ 
  APOL6 & 0.091 (0.022, 0.16) & 0.042 & 1.2e-06 & 1.2e-02 \\ 
  BID & 0.091 (0.022, 0.16) & 0.042 & 1.2e-06 & 1.2e-02 \\ 
  CASP5 & 0.091 (0.022, 0.16) & 0.042 & 1.2e-06 & 1.2e-02 \\ 
  CASZ1 & 0.091 (0.022, 0.16) & 0.042 & 1.2e-06 & 1.2e-02 \\ 
  CD40 & 0.091 (0.022, 0.16) & 0.042 & 1.2e-06 & 1.2e-02 \\ 
  ETV7 & 0.091 (0.022, 0.16) & 0.042 & 1.2e-06 & 1.2e-02 \\ 
  IL18BP & 0.091 (0.022, 0.16) & 0.042 & 1.2e-06 & 1.2e-02 \\ 
  KCNJ15 & 0.091 (0.022, 0.16) & 0.042 & 1.2e-06 & 1.2e-02 \\ 
  LAMP3 & 0.091 (0.022, 0.16) & 0.042 & 1.2e-06 & 1.2e-02 \\ 
  LGALS3BP & 0.091 (0.022, 0.16) & 0.042 & 1.2e-06 & 1.2e-02 \\ 
  LMO2 & 0.091 (0.022, 0.16) & 0.042 & 1.2e-06 & 1.2e-02 \\ 
  LTBR & 0.091 (0.022, 0.16) & 0.042 & 1.2e-06 & 1.2e-02 \\ 
  PDK3 & 0.091 (0.022, 0.16) & 0.042 & 1.2e-06 & 1.2e-02 \\ 
  PSMB3 & 0.091 (0.022, 0.16) & 0.042 & 1.2e-06 & 1.2e-02 \\ 
  RTN1 & 0.091 (0.022, 0.16) & 0.042 & 1.2e-06 & 1.2e-02 \\ 
  STAT5A & 0.091 (0.022, 0.16) & 0.042 & 1.2e-06 & 1.2e-02 \\ 
  MARCO & 0.104 (0.039, 0.169) & 0.040 & 1.4e-06 & 1.4e-02 \\ 
  ZFYVE26 & 0.104 (0.039, 0.169) & 0.040 & 1.4e-06 & 1.4e-02 \\ 
  BTN3A1 & 0.117 (0.056, 0.178) & 0.037 & 1.4e-06 & 1.5e-02 \\ 
\bottomrule
\end{longtable}

\begin{table*}[ht]%
\centering %
\caption{Sensitivity analysis evaluating the effect of varying the non-inferiority margin $\epsilon$, where values closer to 0 result in fewer candidate surrogates to combine for the evaluation stage. The evaluation metric for the combined marker, $\boldsymbol{\delta_{\gamma_{S}}}$, its standard deviation, and its p-value corresponding to a test based on a desired power of 90\% are given in the table.
\label{tables2}}%
\begin{tabular*}{\textwidth}{@{\extracolsep\fill}lllll@{\extracolsep\fill}}
\toprule
$\boldsymbol{\epsilon}$  \textbf{(screening)}  & \textbf{No. of genes in} $\boldsymbol{\gamma_{S}}$ & $\boldsymbol{\delta_{\gamma_{S}}}$ \textbf{(95\% C.I.)} & $\boldsymbol{\sigma_{\delta_{\gamma_{S}}}}$ & \textbf{p-value} \\ 
\midrule
  0.05  &  0 &  &  &  \\  
  0.1  &  0 &  &  & \\  
  0.15 &  64 & -0.038 (-0.102, 0.025) & 0.038 & 3.1e-03 \\ 
  0.20 & 117 & -0.038 (-0.102, 0.025) & 0.038 & 3.1e-03 \\ 
  0.25 & 165 & -0.038 (-0.102, 0.025) & 0.038 & 3.1e-03 \\ 
  0.30 & 246 & -0.038 (-0.102, 0.025) & 0.038 & 3.1e-03 \\ 
  0.35 & 301 & -0.038 (-0.102, 0.025) & 0.038 & 3.1e-03 \\ 
\bottomrule
\end{tabular*}
\end{table*}

\begin{table*}[ht]%
\centering %
\caption{Comparison of evaluation results between the composite surrogate marker \(\gamma_{\mathcal{S}}\), constructed from the 222 significant genes identified in the screening stage, and the top 10 genes from the screening stage evaluated individually in the evaluation data.\label{tables3}}%
\begin{tabular*}{\textwidth}{@{\extracolsep\fill}llll@{\extracolsep\fill}}
\toprule
\textbf{Marker} & $\boldsymbol{\delta}$ & $\boldsymbol{\sigma}$ & \textbf{p-value} \\ 
\midrule
$\gamma_{\mathcal{S}}$ & -0.038 (-0.102, 0.025) & 0.038 & 3.1e-03 \\ 
  CNDP2 & 0 (-0.091, 0.091) & 0.055 & 4.8e-03 \\ 
  IFI44L & -0.038 (-0.102, 0.025) & 0.038 & 3.1e-03 \\ 
  IFITM3 & -0.038 (-0.102, 0.025) & 0.038 & 3.1e-03 \\ 
  NPC2 & -0.038 (-0.102, 0.025) & 0.038 & 3.1e-03 \\ 
  PSME1 & -0.038 (-0.102, 0.025) & 0.038 & 3.1e-03 \\ 
  SERPING1 & -0.038 (-0.102, 0.025) & 0.038 & 3.1e-03 \\ 
  VAMP5 & -0.038 (-0.102, 0.025) & 0.038 & 3.1e-03 \\ 
  EPB41L3 & -0.038 (-0.102, 0.025) & 0.038 & 3.1e-03 \\ 
  IFI6 & -0.038 (-0.102, 0.025) & 0.038 & 3.1e-03 \\
  IRF7 & -0.038 (-0.102, 0.025) & 0.038 & 3.1e-03 \\ 
\bottomrule
\end{tabular*}
\end{table*}

\end{document}